\documentclass[letter, conference]{ieeeconf}  

\IEEEoverridecommandlockouts                              

\title{\LARGE \bf
On Scalable Supervisory Control of Multi-Agent Discrete-Event Systems
}
\usepackage{graphicx}
\usepackage{amsmath,amssymb}
\usepackage{latexsym}
\usepackage{color}
\usepackage{url}
\usepackage{fancyhdr}
\usepackage{setspace}
\usepackage{mathtools}

\def\qed{\hfill $\Box$}
\newtheorem{Theorem}{Theorem}
\newtheorem{Proposition}{Proposition}
\newtheorem{Lemma}{Lemma}

\author{Yingying Liu, Kai Cai, and Zhiwu Li \ \ \ (\today)
\thanks{Yingying Liu is with School of Electro-Mechanical Engineering, Xidian University,
        Xi'an, 710071, China. Kai Cai is with Department of Electrical \& Information Engineering, Osaka City
        University, Osaka, 558-8585, Japan. Zhiwu Li is with the Institute of Systems Engineering, Macau University of Science and Technology, Taipa, Macau, and also with the School of Electro-Mechanical Engineering, Xidian University, Xi'an, 710071, China. {\tt\small Email addresses: liu@c.info.eng.osaka-cu.ac.jp (Y. Liu), kai.cai@eng.osaka-cu.ac.jp (K. Cai), zhwli@xidian.edu.cn (Z. Li)}.}%
}

\begin{document}

\maketitle
\thispagestyle{plain}
\pagestyle{plain}

\begin{abstract}

In this paper we study multi-agent discrete-event systems where the agents can be divided into several groups, and within each group the agents have similar or identical state transition structures. We employ a {\it relabeling} map to generate a "template structure" for each group, and synthesize a {\it scalable} supervisor whose state size and computational process are independent of the number of agents. This scalability allows the supervisor to remain invariant (no recomputation or reconfiguration needed) if and when there are agents removed due to failure or added for increasing productivity. The constant computational effort for synthesizing the scalable supervisor also makes our method  promising for handling large-scale multi-agent systems.
Moreover, based on the scalable supervisor we design scalable local controllers, one for each component agent, to establish a purely distributed control architecture.
Three examples are provided to illustrate our proposed scalable supervisory synthesis and the resulting scalable supervisors as well as local controllers.

\end{abstract}


\section{INTRODUCTION} \label{sec1_intro}

Multi-agent systems have found increasing applications in large-scale engineering practice where tasks are difficult to be accomplished by a single entity. Examples include multiple machines in factories, robots in manufacturing cells, and AGVs in logistic systems \cite{Elm05,WuZhou07,WurDanMou08}. Although not always the case, multi-agent systems typically can be divided into several groups, according to different roles, functions, or capabilities.
For instance, machines are grouped to process different types of workpieces, robots to manufacture different parts of a product, AGVs to transport items of distinct sizes, shapes and weights.
Agents in the same group often have similar or even identical state transition structures, i.e. dynamics. This we shall refer to as a {\it modular} characteristic.

In this paper we study multi-agent systems with such a modular characteristic, and consider individual agents modeled by discrete-event systems (DES). Given a control specification, one may in principle apply supervisory control theory \cite{Ramadge & Wonham (1987a), Ramadge & Wonham (1987),Wonham (2016)} to synthesize a monolithic (i.e. centralized) supervisor for the entire multi-agent system. While the supervisor computed by this method is optimal (i.e. maximally permissive) and nonblocking, there are two main problems. First, the state size of the supervisor increases (exponentially) as the number of agents increases \cite{Gohari & Wonham (2000)}; consequently, the supervisor synthesis will become computationally infeasible for large numbers of agents. Second, whenever the number of agents changes (increases when more agents are added into the system to enhance productivity or to improve redundancy for the sake of reliability; or decreases when some agents malfunction and are removed from the system), the supervisor must be recomputed or reconfigured (e.g. \cite{KumTak12,NooSch15}) in order to adapt to the change.

The first problem may be resolved by decentralized and/or hierarchical supervisory synthesis methods (e.g. \cite{Wong & Wonham (1996), WonLee02, Feng & Wonham (2008), Schmidt & Moor (2008)}). These methods, however, usually can deal only with fixed numbers of agents, and thus must also be recomputed or reconfigured if and when the agent number changes.

In this paper we solve both problems mentioned above by exploiting the modular characteristic of multi-agent systems, and thereby designing a {\it scalable} supervisor whose state number and computational process are {\it independent} of the number of agents. First, owing to similar/identical transition structures of agents in the same group, we employ a {\it relabeling map} (precise definition given in Section II.A below) to generate a ``template structure'' for each group. The template structures thus generated are independent of the agent numbers. Then we design a supervisor based on these template structures, and prove that it is a scalable supervisor for the multi-agent system under certain sufficient condition. The controlled behavior of the designed scalable supervisor need not be optimal, but is nonblocking. Moreover, we show that the sufficient condition for the scalable supervisor is efficiently checkable.

While the designed scalable supervisor serves as a {\it centralized} controller for the multi-agent system,
it may sometimes be natural, and even more desirable, to equip each individual agent with its own {\it local} controller (such that it becomes an autonomous, intelligent agent).
Hence we move on to design {\it scalable} local controllers whose state numbers and computational process are invariant with respect to the number of component agents; for this design,
we employ the method of supervisor localization \cite{Cai & Wonham (2010),Cai & Wonham (2015),Cai & Wonham (2016)}.
Directly localizing the scalable supervisor may be computationally expensive, inasmuch as the localization method requires computing the overall plant model.
To circumvent this problem, we localize the supervisor based on the template structures and thereby derive scalable local controllers without constructing the underlying plant model.
It is proved that the collective controlled behavior of these local controllers is equivalent to that achieved by the scalable supervisor.

The contributions of our work are threefold.
First, our designed centralized supervisor has scalability with respect to the number of agents in the system.
This scalability is a desired feature of a supervisor for multi-agent systems,
inasmuch as it allows the supervisor to remain invariant regardless of how many agents are added to or removed from the system (which may occur frequently due to productivity/reliability concerns or malfunction/repair).
Second, the local controllers we designed for individual agents have the same scalability feature, and are guaranteed to collectively achieve
identical controlled behavior as the centralized supervisor does. With the local controllers `built-in', the agents become autonomous and make their own local decisions;
this is particularly useful in applications like multi-robot systems.
Finally, the computation of the scalable supervisor and local controllers is based solely on template structures and is thus independent of agent numbers as well. As a result, the computation load remains the same even if the number of agents increases; this is advantageous as compared to centralized/decentralized supervisory synthesis methods.


We note that \cite{Eyzell & Cury (2001)} also studied multi-agent systems with a modular characteristic and used group-theoretic tools to characterize symmetry among agents with similar/identical structures. Exploiting symmetry, ``quotient automata'' were constructed to reduce the state size of the composed system, based on which supervisors are synthesized. Quotient automata construction was further employed in \cite{Rohloff & Lafortune (2006)} to develop decentralized synthesis and verification algorithms for multi-agent systems. While the systems considered in \cite{Eyzell & Cury (2001), Rohloff & Lafortune (2006)} are more general than ours in that agents are allowed to share events, the state size of the resulting quotient automata is {\it dependent} on the agent numbers and in the worst case exponential in the number of agents. By contrast, we use the relabeling map approach and synthesize scalable supervisors whose state sizes are independent of agent numbers.


We also note that in \cite{Su (2013)}, an automaton-based modeling framework was presented for multi-agent systems in which the agents' dynamics are instantiated from a finite number of ``templates''; a particular product operation enforcing synchronization on broadcasting or receiving events was proposed to compose the agent dynamics. Building on \cite{Su (2013)}, the work in \cite{Su & Lin (2013)} proposed a method that first decomposes the overall control specification into local ones for individual agents, and then incrementally synthesizes a supervisor based on the local specifications. The presented algorithm for incremental synthesis is (again) dependent on, and in general exponential in, the number of agents.

By extending the ideas in \cite{Su (2013)} and \cite{Su & Lin (2013)},
the work in \cite{Su & Lenna (2017)} proposed a scalable control design for a type of multi-agent systems, 
where an ``agent'' was not just a plant component, but indeed a plant of its own including an imposed specification. 
The ``agents'' were instantiated from a template; for the template, under certain condition, an algorithm was proposed to design a supervisor whose instantiation was shown to work for each ``agent''. 
By contrast, we consider multi-agent systems where each agent is simply a plant component, in particular involving {\it no} specification.
Moreover, the centralized/local scalable supervisors we design are distinct from the supervisor given in \cite{Su & Lenna (2017)}, because our centralized supervisor works effectively for the entire system and local supervisors for individual plant components. 

The work most related to ours is reported in \cite{Jiao & Gan (2015), Jiao (2017)}. Therein the same type of multi-agent systems is investigated and relabeling maps are used to generate template structures. Various properties of the relabeling map are proposed which characterize relations between the relabeled system and the original one. Moreover, a supervisor is designed that is provably independent of agent numbers, when these numbers exceed a certain threshold value. The design of the supervisor is, however, based on first computing the synchronous product of all agents, which can be computationally expensive. This can be relieved by using {\it state tree structures} \cite{Jiao (2017)}, but the computation is still dependent on the agent numbers and thus the supervisor has to be recomputed  or reconfigured whenever the number of agents changes. By contrast, our synthesis is based only on the template structures and thus independent of the agent numbers; furthermore the state size of our designed supervisor is {\it always} independent of the number of agents, with no threshold value required.



The rest of this paper is organized as follows. Section~II introduces preliminaries and formulates the scalable supervisory control synthesis problem. Section~III solves the problem by designing a scalable supervisor, and shows that the sufficient condition for solving the problem is efficiently verifiable.
Section~IV designs scalable local controllers for individual agents, and Section~V presents three examples to illustrate scalable supervisors and local controllers.
Finally Section~VI states our conclusions.


\section{Preliminaries and Problem Formulation} \label{sec2_probm}

\subsection{Preliminaries}

Let the DES plant to be controlled be modeled by a {\it generator}
\begin{center}
$\textbf{G}=(Q,\Sigma,\delta,q_0,Q_m)$
\end{center}
where $\Sigma=\Sigma_c\dot{\cup} \Sigma_u$ is a finite event set that is partitioned into a controllable event subset and an uncontrollable subset, $Q$ is the finite state set, $q_0\in Q$ the initial state, $Q_m\subseteq Q$ the set of marker states, and $\delta:Q\times \Sigma\rightarrow Q$ the (partial) transition function. Extend $\delta$ in the usual way such that $\delta:Q\times \Sigma^{*}\rightarrow Q$. The {\it closed behavior} of \textbf{G} is the language
\begin{center}
$L(\textbf{G}):=\{s\in \Sigma^{*}\mid \delta(q_0,s)!\}\subseteq \Sigma^{*}$
\end{center}
in which the notation $\delta(q_0,s)!$ means that $\delta(q_0,s)$ is defined. The {\it marked behavior} of \textbf{G} is
\begin{center}
$L_m(\textbf{G}):=\{s\in L(\textbf{G})\mid \delta(q_0,s)\in Q_m\}\subseteq L(\textbf{G})$.
\end{center}
 A string $s_1$ is a {\it prefix} of another string \textit{s}, written $s_1\leq s$, if there exists $s_2$ such that $s_1s_2$ = $s$.  The {\it prefix closure} of $L_m(\textbf{G})$ is
 \begin{center}
   $\overline{L_m(\textbf{G})}:= \{s_1\in\Sigma^{*} \mid (\exists s\in L_m(\textbf{G})) s_1\leq s\}$.
 \end{center}
We say that \textbf{G} is \emph{nonblocking} if $\overline{L_m(\textbf{G})}= L(\textbf{G})$.

A language $K\subseteq L_m(\textbf{G})$ is \textit{controllable} with respect to $L(\textbf{G})$ provided $ \overline{K}\Sigma_u \cap L(\textbf{G}) \subseteq \overline{K}$ \cite{Wonham (2016)}. Let $E\subseteq L_m({\bf G})$ be a specification language for \textbf{G}, and define the set of all sublaguages of \textit{E} that are controllable with respect to L(\textbf{G}) by
\begin{align*}
\mathcal{C}(E) := \{K\subseteq E \mid \overline{K}\Sigma_u \cap L(\textbf{G}) \subseteq \overline{K} \}.
\end{align*}
Then $\mathcal{C}(E)$ has a unique supremal element \cite{Wonham (2016)}
\begin{align*}
\sup\mathcal{C}(E)=\cup \{K|K\in \mathcal{C}(E)\}.
\end{align*}


For describing a modular structure of plant \textbf{G}, we first introduce a relabeling map.
Let $T$ be a set of new events, i.e. $\Sigma \cap T=\emptyset$. Define a \textit{relabeling} map $R: \Sigma\rightarrow T$ such that for every $\sigma\in\Sigma$,
\begin{align*}
R(\sigma) = \tau, \ \ \ \tau \in T.
\end{align*}
In general $R$ is surjective but need not be injective.

For $\sigma\in\Sigma$, let $[\sigma]$ be the set of events in $\Sigma$ that have the same $R$-image as $\sigma$, i.e.
\begin{align*}
[\sigma] := \{\sigma'\in \Sigma | R(\sigma')=R(\sigma)\}.
\end{align*}
Then $\Sigma = [\sigma_1]\dot{\cup}[\sigma_2]\dot{\cup}\cdots\dot{\cup}[\sigma_k]$, for some $k\geq 1$, and $T$ can be written as $T=\{R(\sigma_1),R(\sigma_2),\ldots,R(\sigma_k)\}$.
We require that $R$ preserve controllable/uncontrollable status of events in $\Sigma$; namely $R(\sigma)$ is a controllable event if and only if $\sigma\in\Sigma_c$.
Thus $T_c:=\{R(\sigma)| \sigma \in \Sigma_c\}$, $T_u:=\{R(\sigma)| \sigma \in \Sigma_u\}$, and $T=T_c \dot{\cup} T_u$.

We extend $R$ such that $R: \Sigma^{*}\rightarrow T^{*}$ according to

(i) $R(\varepsilon) = \varepsilon$, where $\varepsilon$ denotes the empty string;

(ii) $R(\sigma) = \tau$, $\sigma \in \Sigma$ and $\tau \in T$;

(ii) $R(s\sigma) = R(s)R(\sigma)$, $\sigma\in\Sigma$ and $s\in \Sigma^*$.

\noindent Note that $R(s) \neq \varepsilon$ for all $s \in \Sigma^* \setminus \{\varepsilon\}$.

Further extend $R$ for languages, i.e. $R:Pwr(\Sigma^{*})\rightarrow Pwr(T^{*})$, and define
\begin{align*}
R(L)=\{R(s) \in T^* | s\in L\},\ \ L\subseteq \Sigma^{*}.
\end{align*}
The \textit{inverse-image function} $R^{-1}$ of $R$ is given by $R^{-1}:$ $Pwr(T^{*})\rightarrow Pwr(\Sigma^{*})$:
\begin{center}
 $R^{-1}(H)=\{s\in \Sigma^{*}| R(s)\in H\}$, \ $H\subseteq T^{*}$.
\end{center}
Note that $RR^{-1}(H)=H$, $H\subseteq T^{*}$
while $R^{-1}R(L)\supseteq L$, $L\subseteq \Sigma^{*}$.
We say that $L \subseteq \Sigma^*$ is {\it (${\bf G}, R$)-normal} if $R^{-1}R(L) \cap L_m({\bf G}) \subseteq L$; this property will turn out to be important in Section~III below. Several useful properties of $R$ and $R^{-1}$ are presented in the following lemma, whose proof is given in Appendix.

\begin{Lemma} \label{lem:cr}
For $R:Pwr(\Sigma^{*})\rightarrow Pwr(T^{*})$ and $R^{-1}:$ $Pwr(T^{*})\rightarrow Pwr(\Sigma^{*})$, the following statements are true.

(i) $R(\overline{L})= \overline{R(L)}$, $L\subseteq \Sigma^{*}$.

(ii) $R(L_1 \cap L_2)\subseteq R(L_1)\cap R(L_2)$, $L_1, L_2\subseteq \Sigma^{*}$.

(iii) $R^{-1}(\overline{H})= \overline{R^{-1}(H)}$, $H\subseteq T^{*}$.

(iv) $R^{-1}(H_1 \cap H_2)= R^{-1}(H_1)\cap R^{-1}(H_2)$, $H_1, H_2\subseteq T^{*}$.
\end{Lemma}

\smallskip

We now discuss computation of $R$, $R^{-1}$ by generators.
Let $R: \Sigma^{*}\rightarrow T^{*}$ be a relabeling map and $\textbf{G}=(Q,\Sigma,\delta,q_0,Q_m)$ a generator.  First, relabel each transition of {\bf G} to obtain ${\bf G}_T = (Q,T,\delta_T,q_0,Q_m)$, where $\delta_T : Q \times T \rightarrow Q$ is defined by
\begin{align*}
\delta_T(q_1, \tau) = q_2 \mbox{ iff } (\exists \sigma \in \Sigma) R(\sigma)=\tau \ \&\ \delta(q_1,\sigma)=q_2.
\end{align*}
Hence $L_m(\textbf{G}_T)=R(L_m(\textbf{G}))$ and $L(\textbf{G}_T)=R(L(\textbf{G}))$. However, ${\bf G}_T$ as given above may be {\it nondeterministic} \cite{Wonham (2016)}. Thus apply {\it subset construction} \cite{Wonham (2016)} to convert ${\bf G}_T$ into a deterministic generator $\textbf{H}=(Z,T,\zeta,z_0,Z_m)$, with $L_m(\textbf{H})=L_m(\textbf{G}_T)$ and $L(\textbf{H})=L(\textbf{G}_T)$.\footnote{The worst-case complexity of subset construction is exponential. In the problem considered in this paper, nevertheless, the generators that need to be relabeled typically have small state sizes, and hence their relabeled models may be easily computed. This point will be illustrated by examples given below.}
See Fig.~\ref{fig:relabel_generator} for an illustrative example.

\begin{Lemma} \label{lem:nonb}
If $\textbf{G}$ is nonblocking, then the relabeled generator $\textbf{H}$ is also nonblocking.
\end{Lemma}
{\it Proof.} Suppose that $\textbf{G}$ is nonblocking, i.e. $\overline{L_m(\textbf{G})}=L(\textbf{G})$. Then
\begin{align*}
&R(\overline{L_m(\textbf{G})})=R(L(\textbf{G}))\\
\Rightarrow &\overline{R(L_m(\textbf{G}))}=R(L(\textbf{G})) \ \ \mbox{ (by Lemma~\ref{lem:cr}(i))}\\
\Rightarrow &\overline{L_m(\textbf{H})}=L(\textbf{H})
\end{align*}
namely ${\bf H}$ is nonblocking. \qed

Conversely, to inverse-relabel {\bf H}, simply replace each transition $\tau (\in T)$ of {\bf H} by those $\sigma (\in \Sigma)$ with $R(\sigma)=\tau$;  thus one obtains ${\bf G}'=(Z,\Sigma,\zeta',z_0,Z_m)$, where $\zeta' : Z \times \Sigma \rightarrow Z$ is defined by
\begin{align*}
\zeta'(z_1, \sigma) = z_2 \mbox{ iff } (\exists \tau \in T) R(\sigma)=\tau \ \&\ \zeta(z_1,\tau)=z_2.
\end{align*}
It is easily verified that $L_m(\textbf{G}')=R^{-1}L_m(\textbf{H})$ and $L(\textbf{G}')=R^{-1}L(\textbf{H})$.
Note that ${\bf G}'$ as given above is deterministic (since {\bf H} is), and has the same number of states as {\bf H}; namely inverse-relabeling does not change state numbers. Note that $L_m(\textbf{G}') \supseteq L_m(\textbf{G})$ and $L(\textbf{G}') \supseteq L(\textbf{G})$. Refer again to Fig.~\ref{fig:relabel_generator} for illustration.
Henceforth we shall write $R(\textbf{G}) := \textbf{H}$ and $R^{-1}({\bf H}) := {\bf G}'$.

\begin{figure}
  \centering
  \includegraphics[width=0.48\textwidth]{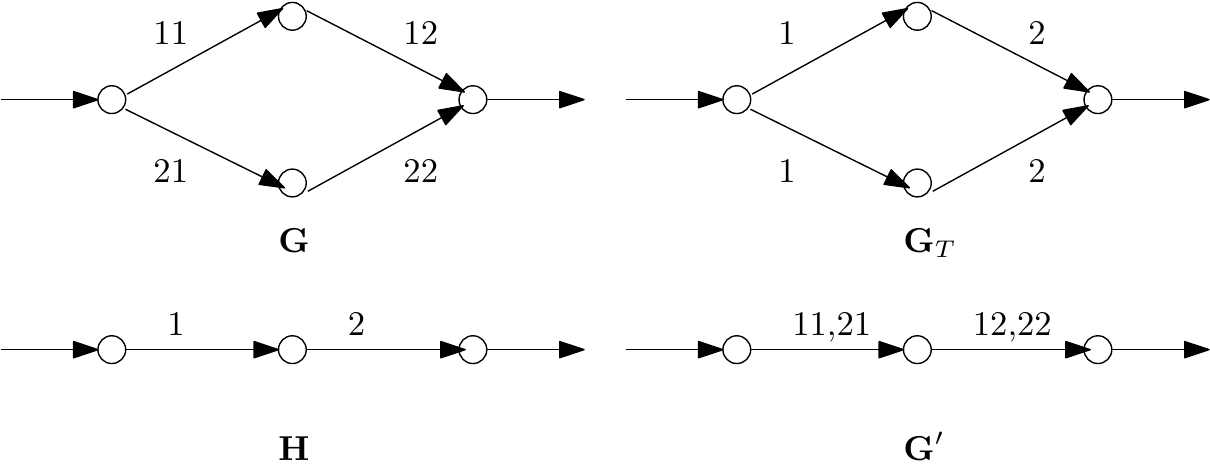}
  \caption{Consider the generator \textbf{G} as displayed and a relabeling map $R: \Sigma^{*}\rightarrow T^{*}$ with $\Sigma=\{11,21,12,22\}$, $T=\{1,2\}$, $R(11)=R(21)=1$ and $R(12)=R(22)=2$. First, relabel each transition of {\bf G} to obtain $\textbf{G}_T$. Evidently $\textbf{G}_T$ is nondeterministic. Thus apply subset construction on ${\bf G}_T$ to derive a deterministic generator $\textbf{H}$. It is easily checked that $L_m(\textbf{H})=R(L_m(\textbf{G}))$ and $L(\textbf{H})=R(L(\textbf{G}))$. To inverse-relabel $\textbf{H}$, replace transition 1 by 11,21 and 2 by 12, 22; thereby one obtains the generator $\textbf{G}'$. It is verified that $L_m(\textbf{G}')=R^{-1}(L_m(\textbf{H}))$ and $L(\textbf{G}')=R^{-1}(L(\textbf{H}))$. Note that $\textbf{G}'$ and \textbf{H} have the same number of states. Convention: the initial state of a generator is labeled by a circle with an entering arrow, while a marker state is labeled by a circle with an exiting arrow. The same notation will be used in subsequent figures.}
  \label{fig:relabel_generator}
\end{figure}

\subsection{Problem Formulation}
Let $R: \Sigma^{*}\rightarrow T^{*}$ be a relabeling map, and $\mathcal{G}=\{\textbf{G}_{1},\ldots,\textbf{G}_{k}\}$ be a set of generators. We say that $\mathcal{G}$ is a \textit{similar set} under $R$ if there is a generator \textbf{H} such that
\begin{align} \label{eq:similarset}
(\forall i\in \{1,\ldots,k\}) R(\textbf{G}_i)=\textbf{H}.
\end{align}
One may view {\bf H} as a ``template'' for $\mathcal{G}$ in that each generator ${\bf G}_i$ in the set may be relabeled to {\bf H}.\footnote{More generally, one may consider {\it DES isomorphism} (e.g. \cite{Cai & Wonham (2010)}) and say that $\mathcal{G}=\{\textbf{G}_{1},\ldots,\textbf{G}_{k}\}$ is a similar set if $R(\textbf{G}_i)$ and $R(\textbf{G}_j)$ are isomorphic for all $i,j\in \{1,\ldots,k\}$. For simplicity of presentation we use the definition in (\ref{eq:similarset}), and subsequent development may be readily extended to the more general case using DES isomorphism.}

\begin{figure}
  \centering
  \includegraphics[width=0.4\textwidth]{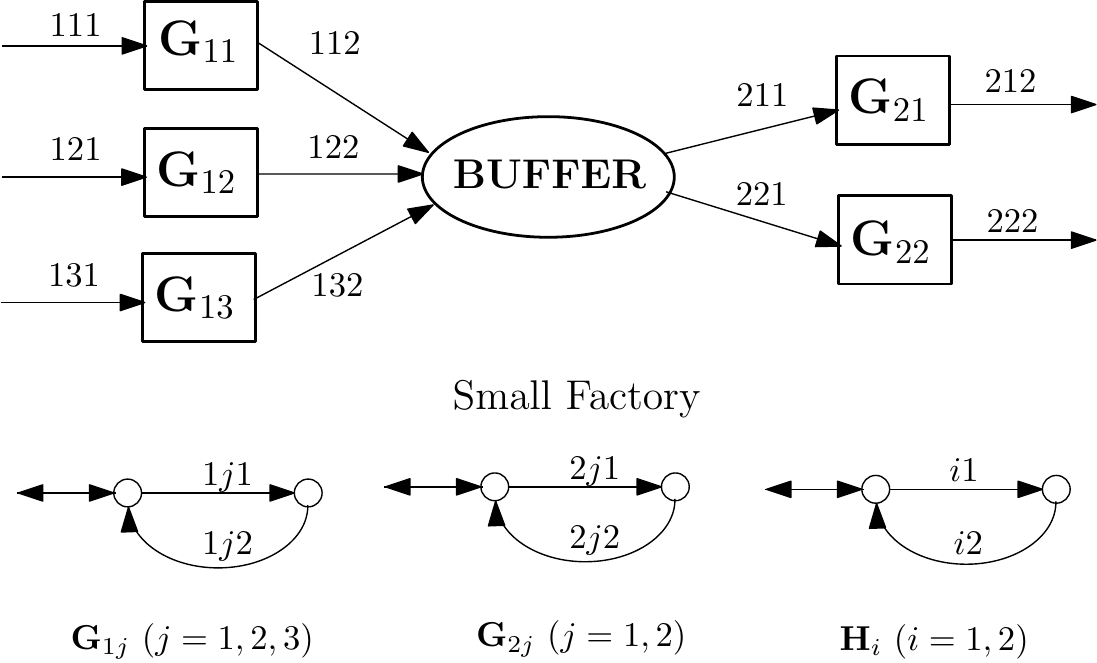}
  \caption{Consider a small factory consisting of  3 input machines ${\bf G_{11}}, {\bf G_{12}}, {\bf G_{13}}$ and 2 output machines ${\bf G_{21}}, {\bf G_{22}}$, linked by a buffer in the middle.  Events $1j1$ ($ j \in \{1,2,3\}$) and $2j1$ ($ j \in \{1,2\}$) mean that machine ${\bf G}_{ij}$ starts to work by taking in a workpiece; events $1j2$ and $2j2$ mean that ${\bf G}_{ij}$ finishes work and outputs a workpiece.
Let $\Sigma = \Sigma_c \dot\cup \Sigma_u = \{111,121,131,211,221\} \dot\cup \{112,122,132,212,222\}$, $T=\{i1,i2 \,|\, i \in \{1,2\}\}$, and a relabeling map $R: \Sigma^* \rightarrow T^*$ with $R(ij1)=i1 \in T_c$, $R(ij2)=i2 \in T_u$ for all $i\in \{1,2\}$. Hence, under $R$, the plant is divided into 2 similar groups  $\{\textbf{G}_{11},\textbf{G}_{12},\textbf{G}_{13}\}$ and $\{\textbf{G}_{21},\textbf{G}_{22}\}$, with template generators $\textbf{H}_1$ and $\textbf{H}_2$ respectively. It is evident that Assumptions (A1) and (A2) hold. Convention: the initial state of a generator is labeled by a circle with an entering arrow, while a marker state is labeled by a circle with an exiting arrow. The same notation will be used in subsequent figures.}
\label{fig:SmallFact22}
\end{figure}

In this paper, the plant {\bf G} is divided into $l (>1)$ groups of component agents, each group $\mathcal{G}_i \,(i \in \{1,\ldots,l\})$ being a similar set of generators under a given relabeling map $R$, i.e. $\mathcal{G}_i = \{ \textbf{G}_{i1},\ldots,\textbf{G}_{i \, n_i} \}$ ($n_i \geq 1$) and there is a generator $\textbf{H}_i$ such that
\begin{align} \label{eq:Hi}
(\forall j \in\{1,\dots,n_i\}) R(\textbf{G}_{ij}) = {\bf H}_i.
\end{align}
Let $\textbf{G}_{ij}$ be defined on $\Sigma_{ij}$ and $\textbf{H}_i$ on $T_i$. Then $R(\Sigma_{ij})=T_i$ for all $j \in \{1,...,n_i\}$.

Note that we do not consider the case where {\bf G} is divided into only one group (i.e. $l=1$),
because the control specifications considered in this paper are imposed {\it between} different groups.
Also we shall demonstrate in Section~V.C how to transform the problem where {\bf G} (naturally) contains only one group of agents into our setup.

Now we make the following assumptions.

\noindent (A1) All component agents are nonblocking and independent, i.e. their event sets are pairwise disjoint.\footnote{Under (A1), ${\bf H}_i$ ($i \in \{1,\ldots,l\}$) computed from (\ref{eq:Hi}) are nonblocking by Lemma~\ref{lem:nonb}.}


\noindent (A2) The template generators ${\bf H}_i$ ($i \in \{1,\ldots,l\}$) have pairwise-disjoint event sets. (This assumption can be regarded as being imposed on the relabeling map $R$, since the event set $T_i$ of ${\bf H}_i$ is obtained by relabeling those $\Sigma_{ij}$ of $\textbf{G}_{ij}$, $j \in \{1,\ldots,n_i\}$.)

As described above, the plant {\bf G} represents a multi-agent DES with a {\it modular} structure, i.e. containing multiple groups of similar and independent agents.
Although it would be more general to consider event sharing among agents, this modular structure is not uncommon in practical multi-agent systems
(e.g. machines in factories, robots in warehouses, and vehicles at intersections). One example of this type of modular plant is given in Fig.~\ref{fig:SmallFact22}; more examples will be illustrated in Section~V below.

%
%
%
%
%
%

Let $\Sigma$ ($=\Sigma_c \dot\cup \Sigma_u$) be the event set of plant ${\bf G}$, and $E \subseteq \Sigma^*$ a specification language that imposes behavioral constraints on ${\bf G}$ (thus the specification with respect to the plant is $E \cap L_m(\textbf{G})$).
 We make the following assumption on the specification.

(A3) The specification language $E$ can be represented by a (nonblocking) generator ${\bf E}$ (i.e. $L_m({\bf E}) = E$) that satisfies $R^{-1}(R({\bf E}))= {\bf E}$. 

This assumption implies that $E$ is $({\bf G}, R)$-normal, i.e. $R^{-1}R(E)\cap L_m(G) \subseteq E$. 
To check if (A3) holds, first compute $R^{-1}(R({\bf E}))$ as described in Section 2.1, and then verify if the result is DES isomorphic (e.g. [1]) to ${\bf E}$. 

Now with plant ${\bf G}$ and specification $E$, the standard supervisory control design \cite{Wonham (2016)} proceeds as follows. First compute the plant ${\bf G}$ by {\it synchronous product} \cite{Wonham (2016)} of all component agents:
\begin{align*}
{\bf G} = ||_{i \in \{1,\ldots,l\}} {\bf G}_i,\ \mbox{ where } {\bf G}_{i} = ||_{j \in \{1,\ldots,n_i\}} {\bf G}_{ij}.
\end{align*}
Under Assumption (A1), {\bf G} is nonblocking.
Then synthesize a supervisor {\bf SUP} (a nonblocking generator) such that\footnote{A supervisor is formally defined as a map associating each string in the closed behavior of {\bf G} with a {\it control patter}, i.e. a subset of enabled events. The generator supervisor {\bf SUP} we use is an {\it implementation} of such a map.}
\begin{align*}
L_m({\bf SUP}) = \sup\mathcal{C}(E \cap L_m(\textbf{G})).
\end{align*}
To rule out the trivial case, we assume the following.

\smallskip
\noindent (A4) $L_m({\bf SUP}) \neq \emptyset$ for $n_i = 1$, $i \in \{1,\ldots,l\}$.  Denote this special {\bf SUP} by {\bf SUP1} henceforth, which is the monolithic supervisor when plant {\bf G} contains exactly one agent in each group.
\smallskip


By this synthesis method, the number of states of {\bf SUP} increases (exponentially) as the number of agents ($n_i$, $i \in \{1,\ldots,l\}$) increases, and consequently the supervisor synthesis becomes computationally difficult (if not impossible). In addition, whenever the number $n_i$ of agents changes (e.g. an operating agent malfunctions and is removed from the system, or a new agent/machine is added to increase productivity), the supervisor {\bf SUP} has to be recomputed or reconfigured.

These two problems may be resolved if one can synthesize a supervisor whose state size, as well as the computational effort involved in its synthesis, is {\it independent} of the number $n_i$ of agents, by exploiting the modular structure of the plant ${\bf G}$. We will call such a supervisor {\it scalable}, where scalability is with respect to the number of agents in the plant.

With this motivation, we formulate the following Scalable Supervisory Control Synthesis Problem (SSCSP):

\smallskip
{\it
Design a scalable supervisor {\bf SSUP} (a nonblocking generator) such that

\noindent (i) The number of states of {\bf SSUP} and its computation are independent of the number $n_i$ of agents for all $i \in \{1,\ldots,l\}$;

\noindent (ii) $L_m({\bf SSUP}) \cap L_m({\bf G})$ satisfies $L_m({\bf SUP1}) \subseteq$ $L_m({\bf SSUP}) \cap L_m({\bf G}) \subseteq L_m({\bf SUP})$.
}

Condition (ii) requires that $L_m({\bf SSUP}) \cap L_m({\bf G})$ be controllable with respect to $L({\bf G})$, and be lower-bounded by the marked behavior of {\bf SUP1}.
It would be ideal to have $L_m({\bf SSUP}) \cap L_m({\bf G}) = L_m({\bf SUP})$. Inasmuch as this requirement might be too strong to admit any solution to the problem, we shall consider (ii) above.
\smallskip

\section{Scalable Supervisory Control}

In this section we design a scalable supervisor to solve the Scalable Supervisory Control Synthesis Problem (SSCSP), under  a easily-verifiable condition. 

Consider the plant {\bf G} as described in Section~II.B. Let $\Sigma (=\Sigma_c \dot\cup \Sigma_u)$ be the event set of {\bf G}, and $R : \Sigma \rightarrow T$ a relabeling map. The procedure of designing a scalable supervisor is as follows, (P1)-(P4), which involves first synthesizing a supervisor for `relabeled system' under $R$ and then inverse-relabeling the supervisor.

\smallskip

\noindent (P1)  Let $k_i \in \{1,...,n_i\}$ denote the number of agents in group $i$ allowed to work in parallel, and compute ${\bf M}_i := R(||_{j=1,\dots,k_i} {\bf G}_{ij})$. Then compute the relabeled plant ${\bf M}$ as the synchronous product of the generators ${\bf M}_i$, i.e.
\begin{align} \label{eq:M}
{\bf M} := ||_{i \in \{1,...,l\}} {\bf M}_i.
\end{align}
We call ${\bf M}$ the {\it relabeled plant} under $R$; it is nonblocking by Assumptions (A1), (A2). The event set of ${\bf M}$ is $T =T_c \dot\cup T_u$, where $T_c = R(\Sigma_c)$ and $T_u=R(\Sigma_u)$.  For computational trackability, one would choose $k_i$ to be (much) smaller than $n_i$. When all $k_i=1$, we have the special case addressed in \cite{Yingying(2018)}. Note that once $k_i$ are fixed, the state sizes of ${\bf M}_i$ and ${\bf M}$ are fixed as well, and independent of the number $n_i$ of agents in group $i$.

\noindent (P2) Compute $F := R(E)$, where $E \subseteq \Sigma^*$ is the specification imposed on {\bf G}. We call $F \subseteq T^*$ the {\it relabeled specification} imposed on {\bf H}.

\noindent (P3) Synthesize a {\it relabeled supervisor} {\bf RSUP} (a nonblocking generator) such that
\begin{align*}
L_m({\bf RSUP}) = \sup\mathcal{C}(L_m(\textbf{H}) \cap F) \subseteq T^*.
\end{align*}
The number of states of {\bf RSUP} is independent of the number of agents, since ${\bf H}$'s state size is so.

\noindent (P4) Inverse-relabel {\bf RSUP} to derive {\bf SSUP}, i.e.
\begin{align}\label{eq:SSUP}
{\bf SSUP} := R^{-1} ({\bf RSUP})
\end{align}
with the marked behavior
\begin{align*}
L_m({\bf SSUP}) = R^{-1} L_m({\bf RSUP}) \subseteq \Sigma^*.
\end{align*}

\begin{figure}
  \centering
  \includegraphics[width=0.48\textwidth]{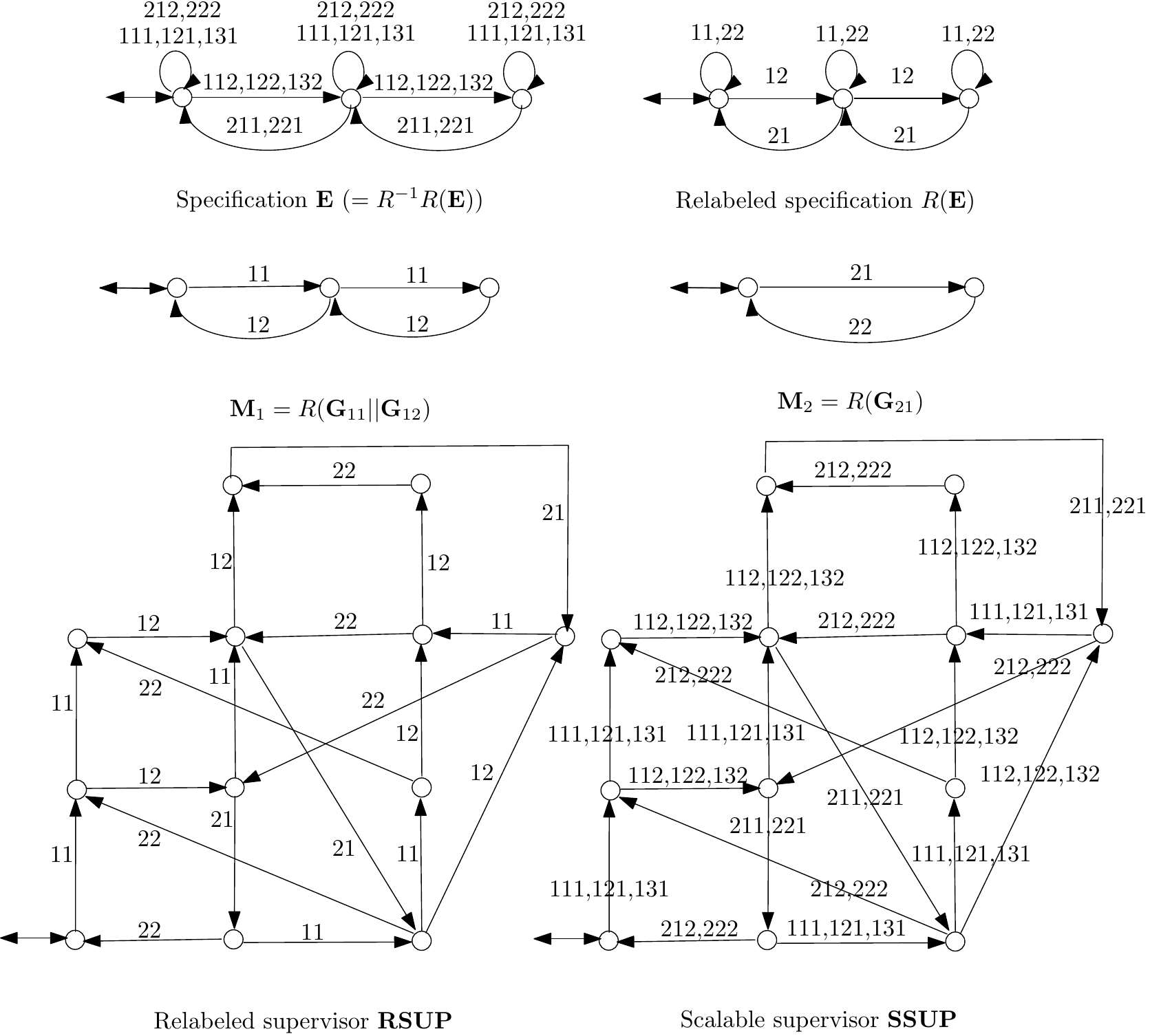}\\
  \caption{Consider the small factory example in Fig.~\ref{fig:SmallFact22}, and a specification that protects the buffer (with two slots) against overflow and underflow. This specification is represented by {\bf E},  which satisfies (A3). In (P1), compute the relabeled plant ${\bf M} = {\bf M}_1 \| {\bf M}_2$, where ${\bf M}_1=R(\textbf{G}_{11}|| \textbf{G}_{12}),\ {\bf M}_2=R(\textbf{G}_{21})$ for $k_1=2$ and $k_2=1$.  In (P2), compute the relabeled specification $R({\bf E})$. In (P3), compute the relabeled supervisor {\bf RSUP} for {\bf M} and $R({\bf E})$. Finally in (P4), compute $R^{-1}({\bf RSUP})$ to derive the scalable supervisor {\bf SSUP}. Note that the relabeled plant {\bf M} and  the scalable supervisor {\bf SSUP} allow at most two machines in the input group to work in parallel as $k_1=2$. }
  \label{fig:SmallFact22_ScaSup}
\end{figure}

By the inverse-relabeling computation introduced in Section~II.A, {\bf SSUP} computed in (\ref{eq:SSUP}) has the same number of states as {\bf RSUP}. It then follows that the state size of {\bf SSUP} is independent of the number of agents in plant {\bf G}.\footnote{Note that the state size of {\bf SSUP} is related to the number of groups that the plant is divided into, as well as the state size of the generator representing the relabeled specification $F$. In this paper we focus on the scalability of supervisor with respect to the number of agents, and thus assume the above two factors fixed for each problem we consider. In applications where these factors may be relevant, different approaches will need to be developed.}  Moreover, it is easily observed that {\bf SSUP} is nonblocking (since {\bf RSUP} is), and its computation does not depend on the number $n_i$ of agents in each group $i$ ($\in \{1,\ldots,l\}$). The above design procedure is demonstrated with an example displayed in Fig.~\ref{fig:SmallFact22_ScaSup}.

\footnotetext[6] The scalable supervisor {\bf SSUP} has the same number of states and structure as {\bf SUP1} (the monolithic supervisor when plant {\bf G} contains exactly one agent in each group). Note, however, that {\bf SSUP} is more permissive than
{\bf SUP1} (having strictly larger closed and marked behaviors), and is more flexible in that {\bf SSUP} imposes {\it no} restriction on the order by which the agents in the same group (input machines or output machines).

Our main result is the following.

\begin{Theorem} \label{thm:main}
Consider the plant {\bf G} as described in Section~II.B and suppose that Assumptions (A1), (A2), (A3), and (A4) hold.
If $L_m(\textbf{M})$ is controllable with respect to $R(L(\textbf{G}))$, then {\bf SSUP} in (\ref{eq:SSUP}) is a scalable supervisor that solves SSCSP.
\end{Theorem}

Theorem~\ref{thm:main} provides a sufficient condition under which {\bf SSUP} in (\ref{eq:SSUP}) is a solution to SSCSP. This condition is the controllability of $L_m({\bf H})$ with respect to $R(L({\bf G}))$, i.e.
$\overline{L_m({\bf H})} \Sigma_u \cap R(L({\bf G})) \subseteq \overline{L_m({\bf H})}$.
This means that the relabeled plant should be controllable with respect to the relabeling of the original plant {\bf G}; in other words, the relabeling operation should not remove uncontrollable events that are allowed by {\bf G}. As we shall see below, this condition is essential in proving the controllability of $L_m({\bf SSUP}) \cap L_m({\bf G})$ with respect to $L({\bf G})$.

For the success of our scalable supervisory control synthesis, it is important to be able to efficiently verify  this sufficient condition. At the appearance, however, this condition seems to require computing {\bf G} which would be computationally infeasible for large systems.
Nevertheless, we have the following result.

\begin{Proposition} \label{prop:check}
Consider the plant {\bf G} as described in Section~II.B and suppose that Assumptions (A1), (A2) hold.
  For each group $i \in \{1,\ldots,l\}$ if $L_m(\textbf{H}_{i})$ is controllable with respect to $R(L(\textbf{G}_{i1}\|\textbf{G}_{i2}))$, then $L_m(\textbf{M})$ is controllable with respect to $R(L(\textbf{G}))$.
\end{Proposition}

Proposition~\ref{prop:check} asserts that the controllability of $L_m(\textbf{M})$ with respect to $R(L(\textbf{G}))$ may be checked in a modular fashion: namely it is sufficient to check the controllability of $L_m(\textbf{H}_{i})$ for each group with respect to only two component agents. As a result, the computational effort of checking the condition is low.

\begin{figure}
  \centering
  \includegraphics[width=0.4\textwidth]{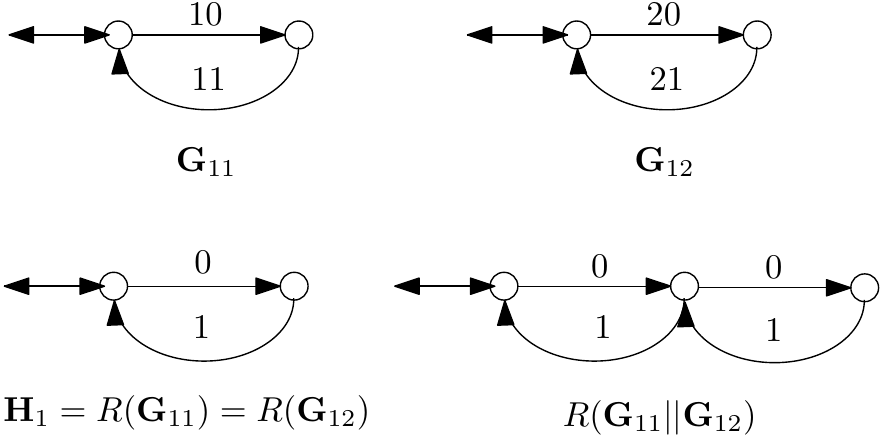}\\
  \caption{Consider a group of 2 machines ${\bf G_{11}}, {\bf G_{12}}$.  Let $\Sigma = \Sigma_c \dot\cup \Sigma_u$, where $\Sigma_c=\{11,21\}$ and $\Sigma_u=\{10,20\}$. Let $T=\{0,1\}$, and the relabeling map $R: \Sigma \rightarrow T$ with $R(11)=R(21)=1 \in T_c$, $R(10)=R(20)=0 \in T_u$. Under $R$, $\textbf{H}_1=R({\bf G}_{11})=R({\bf G}_{12})$ and $R(\textbf{G}_{11}\|\textbf{G}_{12})$ are displayed. Observe that $L_m({\bf H}_1)$ is {\it not} controllable with respect to $R(L(\textbf{G}_{11}\|\textbf{G}_{12}))$: let $t=0\in \overline{L_m(\textbf{H}_{1})}$ and $\tau=0\in T_u$ such that $t\tau\in R(L(\textbf{G}_{11}\|\textbf{G}_{12}))$, but $t\tau\notin \overline{L_m(\textbf{H}_{1})}$.}
  \label{fig:RGH}
\end{figure}

Note that the condition in Proposition~\ref{prop:check}, $L_m(\textbf{H}_{i})$ being controllable with respect to $R(L(\textbf{G}_{i1}\|\textbf{G}_{i2}))$, does not always hold. An example where this condition fails is shown in Fig.~\ref{fig:RGH}.

To prove Proposition~\ref{prop:check}, we need the following two lemmas. For convenience it is assumed that Assumptions (A1), (A2) hold henceforth in this subsection.

\begin{Lemma} \label{lem:3.2}
Let $i \in \{1,\ldots,l\}$.
If $L_m(\textbf{H}_{i})=(R(L_m({\bf G}_{i1})))$ is controllable with respect to $R(L(\textbf{G}_{i1}\|\textbf{G}_{i2}))$, then $R(L_m(\textbf{G}_{i1}\|\textbf{G}_{i2}))$ is controllable with respect to $R(L(\textbf{G}_{i1}\|\textbf{G}_{i2}\|\textbf{G}_{i3}))$.
\end{Lemma}

{\it Proof.} Let $i \in \{1,\ldots,l\}$, $t\in \overline{R(L_m(\textbf{G}_{i1}\|\textbf{G}_{i2}))} $, $\tau \in T_u$, and $t\tau \in R(L(\textbf{G}_{i1}\|\textbf{G}_{i2}\|\textbf{G}_{i3}))$. We shall show that $t\tau \in \overline{R(L_m(\textbf{G}_{i1}\|\textbf{G}_{i2}))}$.  By $t\in \overline{R(L_m(\textbf{G}_{i1}\|\textbf{G}_{i2}))}$ we derive
\begin{align*}
&t\in R(\overline{L_m(\textbf{G}_{i1}\|\textbf{G}_{i2})}) = R(L(\textbf{G}_{i1}\|\textbf{G}_{i2})) \\
\Rightarrow &(\exists s \in L(\textbf{G}_{i1}\|\textbf{G}_{i2})) R(s) = t.
\end{align*}
By Assumption (A1), $\textbf{G}_{i1}$, $\textbf{G}_{i2}$, $\textbf{G}_{i3}$ do not share events, the string $s$ can be divided into two cases:

Case 1: $s\in  L(\textbf{G}_{i1})$ (resp. $s\in  L(\textbf{G}_{i2})$). 

Thus for each $\sigma \in \Sigma_u$ with $R(\sigma) = \tau$, if $s\sigma \in L(\textbf{G}_{i1}\|\textbf{G}_{i2}\|\textbf{G}_{i3})$, then
\begin{align*}
&\mbox{either } s\sigma \in L(\textbf{G}_{i1}) \mbox{ if $\sigma$ is an event of $\textbf{G}_{i1}$} \\
&\mbox{or } s\sigma \in L(\textbf{G}_{i1}\|\textbf{G}_{i2}) \mbox{ if $\sigma$ is an event of $\textbf{G}_{i2}$} \\
&\mbox{or } s\sigma \in L(\textbf{G}_{i1}\|\textbf{G}_{i3}) \mbox{ if $\sigma$ is an event of $\textbf{G}_{i3}$}.
\end{align*}
Hence $t\tau = R(s\sigma) \in R(L(\textbf{G}_{i1}\|\textbf{G}_{i2}\|\textbf{G}_{i3}))$ implies
\begin{align*}
&\mbox{either } R(s\sigma) \in R(L(\textbf{G}_{i1})) \\
&\mbox{or } R(s\sigma) \in R(L(\textbf{G}_{i1}\|\textbf{G}_{i2})) \\
&\mbox{or } R(s\sigma) \in R(L(\textbf{G}_{i1}\|\textbf{G}_{i3}))=R(L(\textbf{G}_{i1}\|\textbf{G}_{i2})).
\end{align*}
For this case, $t\tau \in \overline{R(L_m(\textbf{G}_{i1}\|\textbf{G}_{i2}))}$ always hold.

Case 2: $s\notin  L(\textbf{G}_{i1})$ and $s \in L(\textbf{G}_{i1}\|\textbf{G}_{i2})$.

Similarly, for each $\sigma \in \Sigma_u$ with $R(\sigma) = \tau$, if $s\sigma \in L(\textbf{G}_{i1}\|\textbf{G}_{i2}\|\textbf{G}_{i3})$, then
 \begin{align*}
&\mbox{either }  s\sigma \in L(\textbf{G}_{i1}\|\textbf{G}_{i2}) \mbox{ if $\sigma$ is an event of $\textbf{G}_{i2}$ or $\textbf{G}_{i2}$} \\
&\mbox{or } s\sigma \in L(\textbf{G}_{i1}\|\textbf{G}_{i2}\|\textbf{G}_{i3}) \mbox{ if $\sigma$ is an event of $\textbf{G}_{i3}$}.
\end{align*}

We have $t \notin R(L(\textbf{G}_{i1})$ as $s\notin  L(\textbf{G}_{i1})$. For the latter case, 
{\small
\begin{align*}
&t \in R(L(\textbf{G}_{i1}\|\textbf{G}_{i2}),\ t\tau \notin R(L(\textbf{G}_{i1}\|\textbf{G}_{i2})), \ t\tau \in R(L(\textbf{G}_{i1}\|\textbf{G}_{i2}\|\textbf{G}_{i3})).
\end{align*}}
Since $\textbf{G}_{i1}$, $\textbf{G}_{i2}$, and $\textbf{G}_{i3}$ have the same state transition structure and all the events of them are relabeled by the same relabeling map, there must exist string $t'\in T^*$ and $\tau' \in T_u$ such that 
\begin{align*}
&t' \in R(L(\textbf{G}_{i1})),\ t'\tau' \notin R(L(\textbf{G}_{i1})), \ t'\tau' \in R(L(\textbf{G}_{i1}\|\textbf{G}_{i2})).
\end{align*}
However, $R(L_m({\bf G}_{i1}))$ is controllable with respect to $R(L(\textbf{G}_{i1}\|\textbf{G}_{i2}))$. For all $\tau' \in T_u$ if $t' \in R(L(\textbf{G}_{i1})$ and  $t'\tau' \in R(L(\textbf{G}_{i1}\|\textbf{G}_{i2}))$, then $t'\tau' \in R(L(\textbf{G}_{i1})$ must hold, which is conflict with Case 2.

Therefore, after all, $t\tau \in \overline{R(L_m(\textbf{G}_{i1}\|\textbf{G}_{i2}))}$ as required.\qed

\medskip
Applying Lemma~\ref{lem:3.2} inductively, one derives that if $L_m(\textbf{H}_{i})$ is controllable with respect to $R(L(\textbf{G}_{i1}\|\textbf{G}_{i2}))$, then $L_m(\textbf{M}_{i})$ ($i \in \{1,\ldots,l\}$) is controllable with respect to $R(L(||_{j \in \{1,\ldots,k_{i}+1\}} {\bf G}_{ij}))$.

\medskip
\begin{Lemma} \label{lem:3.3}
Let $i \in \{1,\ldots,l\}$.
If $L_m(\textbf{M}_{i})$ ($i \in \{1,\ldots,l\}$) is controllable with respect to $R(L(||_{j \in \{1,\ldots,k_{i}+1\}} {\bf G}_{ij}))$, then $L_m(\textbf{M}_{i})$ is controllable with respect to $R(L(\textbf{G}_{i}))$.
\end{Lemma}

{\it Proof.} Let  $t\in \overline{L_m({\bf M}_i)}$, $\tau \in T_u$, and $t\tau \in R(L(\textbf{G}_{i}))$. We shall show that $t\tau \in \overline{L_m({\bf M}_i)} = L({\bf M}_i)$.  By 
$t\in  \overline{L_m({\bf M}_i)}$ we derive
\begin{align*}
&t\in L({\bf M}_i)= R(L(||_{j \in \{1,\ldots,k_{i}\}} {\bf G}_{ij})) \\
\Rightarrow &(\exists s \in L(||_{j \in \{1,\ldots,k_{i}\}} {\bf G}_{ij})) R(s) = t.
\end{align*}
By Assumption (A1), agents in the same group do not share events, the string $s$ must be in $||_{j \in \{1,\ldots,k_{i}\}} {\bf G}_{ij}$. Thus for each $\sigma \in \Sigma_u$ with $R(\sigma) = \tau$, if $s\sigma \in L(\textbf{G}_{i})$, then
\begin{align*}
&\mbox{either } s\sigma \in L(||_{j \in \{1,\ldots,k_{i}\}} {\bf G}_{ij}) \mbox{ if $\sigma\in \bigcup_{j \in \{1,\ldots,k_{i}\}}\Sigma_{ij}$} \\
&\mbox{or } s\sigma \in L(||_{j \in \{1,\ldots,k_{i}+1\}} {\bf G}_{ij})  \mbox{ if $\sigma$ is an event of $\textbf{G}_{i,k_{i}+1}$}.
\end{align*}
For the former case,  $t\tau = R(s\sigma) \in R(L(||_{j \in \{1,\ldots,k_{i}\}} {\bf G}_{ij}))= L({\bf M}_i)$. 
For the latter case, use the controllability of $L_m(\textbf{M}_{i})$ with respect to $R(L(||_{j \in \{1,\ldots,k_{i}+1\}} {\bf G}_{ij}))$ to derive $R(s\sigma) \in L({\bf M}_i)$. Therefore,  $t\tau \in L({\bf M}_i)$ is proved.\qed

\medskip
We are now ready to present the proof of Proposition~\ref{prop:check}.

\medskip
{\it Proof of Proposition~\ref{prop:check}.}  Let  $t\in \overline{L_m({\bf M})}$, $\tau \in T_u$, and $t\tau \in R(L(\textbf{G}))$. We shall show that $t\tau \in \overline{L_m({\bf M})} = L({\bf M})$.  By 
$t\in  \overline{L_m({\bf M})}$ we derive
\begin{align*}
&t\in L({\bf M})= L(||_{i \in \{1,\ldots,l\}} {\bf M}_{i})=L(\bigcap_{i \in \{1,\ldots,l\}}P_i^{-1}({\bf M}_{i})),\\
\end{align*}
where $P_i:T^*\rightarrow T_i^*.$ We thus get that $P_i(t)\in L({\bf M_i})$. We have $t\tau \in R(L(\textbf{G}))=R(L(||_{i \in \{1,\ldots,l\}} {\bf G}_{i}))=||_{i \in \{1,\ldots,l\}} R(L({\bf G}_{i}))$ (by Assumption (A1)). Hence 
\begin{align*}
&t\tau\in ||_{i \in \{1,\ldots,l\}} R(L({\bf G}_{i}))=\bigcap_{i \in \{1,\ldots,l\}}P_i^{-1}(R(L({\bf G}_{i}))). 
\end{align*}
Hence, $t\tau\in P_i^{-1}(R(L({\bf G}_{i})))$, i.e. $P_i(t\tau)\in R(L({\bf G}_{i}))$.

Combining Lemmas~\ref{lem:3.2} and \ref{lem:3.3}, it directly follows that if $L_m(\textbf{H}_{i})$ ($i \in \{1,\ldots,l\}$) is controllable with respect to $R(L(\textbf{G}_{i1}\|\textbf{G}_{i2}))$, then $L_m(\textbf{M}_{i})$ is controllable with respect to $R(L(\textbf{G}_{i}))$. Therefore, $P_i(t\tau)\in L({\bf M}_{i})$, i.e.  $t\tau \in P_i^{-1}L(({\bf M}_{i}))$. It is derived that $t\tau \in L(||_{i \in \{1,\ldots,l\}} {\bf M}_{i})=L({\bf M})$.
 \qed

Thus under the easily checkable sufficient condition, Theorem~\ref{thm:main} asserts that {\bf SSUP} in (\ref{eq:SSUP}) is a valid scalable supervisor whose state size is independent of the number of agents in the plant. The advantages of this scalability are, (i) computation of {\bf SSUP} is independent of the number of agents and thus this method may handle systems with large numbers of agents; (ii) {\bf SSUP} does not need to be recomputed or reconfigured if and when some agents are removed due to failure or added for increasing productivity.

For the example in Fig.~\ref{fig:SmallFact22_ScaSup}, it is verified that the sufficient condition of Theorem~\ref{thm:main} are satisfied, and therefore the derived scalable supervisor {\bf SSUP} is a solution to SSCSP.

To prove Theorem~\ref{thm:main} we need to the following lemmas.

\begin{Lemma} \label{lem:sub}
Consider the plant {\bf G} as described in Section~II.B and suppose that Assumptions (A1), (A2) hold. Then {\bf H} is nonblocking, and
\begin{center}
$L_m(\textbf{H})\subseteq R(L_m(\textbf{G}))$.
\end{center}
\end{Lemma}

\medskip

\begin{Lemma} \label{lem:3.1}
Consider the plant {\bf G} as described in Section~II.B and suppose that Assumptions (A1), (A2) hold.
Then ${\bf SSUP}$ and ${\bf G}$ are nonconflicting, i.e.
\begin{center}
$\overline{L_m(\textbf{SSUP})\cap L_m(\textbf{G})}= \overline{L_m(\textbf{SSUP})}\cap \overline{L_m(\textbf{G})}$.
\end{center}
\end{Lemma}

\medskip

The proofs of the above Lemmas are referred to Appendix.
Now we are ready to provide the proof of Theorem~\ref{thm:main}.

{\it Proof of Theorem~\ref{thm:main}.}  That the number of states of ${\bf SSUP}$ and its computation are independent of the number $n_i$ of agents for all $i\in \{1,\ldots,l\}$ has been asserted following (P4) of designing {\bf SSUP}. Hence to prove that {\bf SSUP} is a scalable supervisor that solves SSCSP, we will show that  $L_m({\bf SUP1}) \subseteq L_m({\bf SSUP}) \cap L_m({\bf G}) \subseteq L_m({\bf SUP})$.

First we prove that $L_m({\bf SUP1}) \subseteq L_m({\bf SSUP}) \cap L_m({\bf G})$.  Let $s \in L_m({\bf SUP1})$. Then $s \in ||_{i \in [1,l]} L_m({\bf G}_{i1})\subseteq L_m({\bf G})$. Also it is observed from (P1)-(P3) of designing {\bf SSUP} that $R(L_m({\bf SUP1})) = L_m({\bf RSUP})$. Hence $R(s) \in L_m({\bf RSUP})$ and $s \in R^{-1}(L_m({\bf RSUP})) = L_m({\bf SSUP})$. Therefore $s \in L_m({\bf SSUP}) \cap L_m({\bf G})$, and $L_m({\bf SUP1}) \subseteq L_m({\bf SSUP}) \cap L_m({\bf G})$ is proved.

It remains to show that $L_m({\bf SSUP}) \cap L_m({\bf G}) \subseteq L_m({\bf SUP}) = \sup\mathcal{C}(E \cap L_m(\textbf{G}))$.
For this we will prove that (i) $L_m({\bf SSUP}) \cap L_m({\bf G})$ is controllable with respect to $L(\textbf{G})$, and (ii)
$L_m({\bf SSUP}) \cap L_m({\bf G}) \subseteq E \cap L_m({\bf G})$. For (i) let $s \in \overline{L_m({\bf SSUP}) \cap L_m({\bf G})}$, $\sigma\in \Sigma_u$, $s\sigma \in L(\textbf{G})$. Then
\begin{align*}
&s \in \overline{L_m({\bf SSUP}) \cap L_m({\bf G})} \\
\Rightarrow &(\exists t) st \in L_m({\bf SSUP}) \\
\Rightarrow &st \in R^{-1} L_m({\bf RSUP}) \ \ \ \mbox{ (by (P4))}\\
\Rightarrow &R(st) \in L_m({\bf RSUP}) \subseteq L_m({\bf M}) \\
\Rightarrow &R(s) \in \overline{L_m({\bf RSUP})} \ \&\ R(s) \in \overline{L_m({\bf M})}.
\end{align*}
Since $s \sigma \in L({\bf G})$, we have $R(s)R(\sigma) \in R(L({\bf G}))$ where $R(\sigma) \in T_u$ (since $\sigma \in \Sigma_u$). It then follows from the controllability of $L_m({\bf M})$ with respect to $R(L({\bf G}))$ that $R(s)R(\sigma) \in \overline{L_m({\bf M})} = L({\bf M})$ ({\bf M} is nonblocking by Lemma~\ref{lem:sub}). Now use the controllability of $L_m({\bf RSUP})$ with respect to $L({\bf M})$ to derive $R(s)R(\sigma) \in \overline{L_m({\bf RSUP})}$, and in turn
\begin{align*}
&s\sigma \in R^{-1}R(s\sigma) \subseteq R^{-1} \overline{L_m({\bf RSUP})} \\
\Rightarrow &s\sigma \in \overline{R^{-1} L_m({\bf RSUP})} = \overline{L_m({\bf SSUP})}.
\end{align*}
In the derivation above, we have used Lemma~\ref{lem:cr}(iii).
In addition, since $s\sigma \in L({\bf G}) = \overline{L_m({\bf G})}$ ({\bf G} is nonblocking by Assumption (A1)), we have
\begin{align*}
s\sigma \in \overline{L_m({\bf SSUP})} \cap \overline{L_m({\bf G})}
\end{align*}
Under Assumptions~(A1), (A2), it follows from Lemma~\ref{lem:3.1} that {\bf SSUP} and {\bf G} are nonconflicting, i.e. $\overline{L_m({\bf SSUP})} \cap \overline{L_m({\bf G})}= \overline{L_m({\bf SSUP}) \cap L_m({\bf G})}$. Hence $s\sigma\in \overline{L_m({\bf SSUP}) \cap L_m({\bf G})}$, which proves (i).

For (ii) let $s\in L_m({\bf SSUP}) \cap L_m({\bf G})$. Then
\begin{align*}
&s \in R^{-1}L_m({\bf RSUP}) \cap L_m({\bf G}) \\
\Rightarrow &s \in L_m({\bf G}) \ \&\ R(s) \in L_m({\bf RSUP}) \subseteq F =R(E)\\
\Rightarrow &s \in L_m({\bf G}) \ \&\ s \in R^{-1}R(s) \subseteq R^{-1}R(E).
\end{align*}
Since $E$ is (${\bf G}, R$)-normal, i.e. $R^{-1}R(E) \cap L_m({\bf G}) \subseteq E$, we derive $s \in E \cap L_m({\bf G})$, which proves (ii).
The proof is now complete.\qed

From the proof above, note that if $R(L_m({\bf SUP}))\subseteq L_m({\bf RSUP})$, then we derive $L_m({\bf SUP})\subseteq R^{-1}R(L_m({\bf SUP}))\cap L_m({\bf G})\subseteq R^{-1} (L_m({\bf RSUP}))\cap L_m({\bf G}) = L_m({\bf SSUP})\cap L_m({\bf G})$. This leads to the following.

Corollary 1: Consider the plant {\bf G} as described in Section II.B and suppose that Assumptions (A1), (A2), (A3) hold. If the specification $E \subseteq \Sigma^*$ is $({\bf G},R)$-normal, $L_m({\bf M})$ is controllable with respect to $R(L({\bf G}))$, and
$R(L_m({\bf SUP})) \subseteq L_m({\bf RSUP})$, then {\bf SSUP} in (3) is the least restrictive scalable supervisor that solves SSCSP (i.e. $L_m({\bf SSUP})\cap L_m({\bf G}) = L_m({\bf SUP}))$.

Although the least restrictive scalable solution in Corollary~1 is of theoretical interest, the additional condition
$R(L_m({\bf SUP}))\subseteq L_m({\bf RSUP}) $ may be too strong and its verification requires computing the monolithic supervisor {\bf SUP} which itself is infeasible for large multi-agent systems.

 Alternatively, one may explore the threshold of $k_i$ (the number of agents in group $i$ that are allowed to work in parallel) to achieve the least restrictive controlled behavior. For the small factory example in Figs. 1 and 2, the threshold for both $k_1$ and $k_2$ is $2$, the buffer size.  More generally, for a small factory consists of $n_1$ input machines, $n_2$ output machines, and a buffer of size $b\ (\leq n_1, n_2)$, the threshold for both $k_1$ and $k_2$ is $b$. A thorough study on the threshold of $k_i$ that achieves the least restrictive controlled behavior will be pursued in our future work.

Remark 1: In Theorem 1, the condition that $L_m({\bf M})$ is controllable with respect to $R(L({\bf G}))$ rules out the case where agent models start with an uncontrollable event. To address this case, one approach is to replace the relabeled plant {\bf M} in (P1) by ${\bf M} := R({\bf G})$; the rest (P2)-(P4) remain the same. Suppose that the specification $E \subseteq L_m({\bf G})$ is controllable with respect to $L({\bf G})$. Then it is verified that $R(E)$ is controllable with respect to $R(L({\bf G})) = L({\bf M})$ (under Assumptions (A1), (A2), (A3)), (A4). Hence the resulting $L_m({\bf SSUP}) = R^{-1}R(E)$. Therefore, assuming $E$ is $({\bf G},R)$-normal (as in Theorem 1), we derive $L_m({\bf SSUP})\cap L_m({\bf G}) = R^{-1}R(E)\cap L_m({\bf G}) = E = L_m({\bf SUP})$.  The above reasoning leads to the following.

Corollary 2: Consider the plant {\bf G} as described in Section II.B and suppose that Assumptions (A1), (A2), (A3), (A4) hold. Also suppose that the relabeled plant ${\bf M}$ in (P1) is ${\bf M} := R({\bf G})$. If the specification $E\subseteq L_m({\bf G})$ is controllable with respect to $L({\bf G})$, then {\bf SSUP} in (3) solves SSCSP (with $L_m({\bf SSUP})  \cap L_m({\bf G}) = L_m({\bf SUP}))$.

Although Corollary 2 allows agents to start with an uncontrollable event, the assumption that ${\bf M}=R({\bf G})$ requires computing the plant model ${\bf G}$ which is infeasible for large multi-agent systems.  A special case where the conditions in Corollary~2 hold is when the specification $E = L_m({\bf G})$. For the more general case where $E$ is not controllable with respect to $L({\bf G})$, we shall postpone the investigation to our future work.

\section{Extensions for improving permissiveness}

The preceding section presented a synthesis procedure for a scalable supervisor, and provided efficiently checkable condition under which the scalable supervisor is a solution to SSCSP.  In this section, we present an extension to the previous synthesis procedure in order to improve behavioral permissiveness while maintaining scalability. The improved permissiveness comes with the cost of increased computational cost, which demonstrates a tradeoff relation between scalability (state size of supervisor and its computation) and permissiveness.

The extension to improve behavioral permissiveness is to `refine' the relabeling map $R$ in such a way that each group of agents is further divided into subgroups.

Recall from Section II that the relabeling map $R : \Sigma \rightarrow T$ is assumed to satisfy the following. For an agent ${\bf G}_{ij}$ in group $i \in \{1,\ldots, l\}$ (and $j \in \{1,\ldots,n_i\}$), defined over the event set $\Sigma_{ij}$, there holds $R(\Sigma_{ij}) = T_i$; the sets $T_i$ ($i \in \{1,\ldots,l\}$) are pairwise disjoint and $T = \dot\cup_{i \in \{1,\ldots,l\}} T_i$. Now consider the following `refinement' of $R$. Let $T'_i$ be a set disjoint from $T_i$, and define a new relabeling map $R'$ by
\begin{align*}
& R'(\Sigma_{ij}) = T_i,\ j \in [1, \lfloor \frac{n_i}{2} \rfloor]\\
& R'(\Sigma_{ij}) = T'_i,\ j \in [\lfloor \frac{n_i}{2} \rfloor+1, n_i].
\end{align*}
Thus $R'$ relabels the events of the first half agents in each group $i$ to $T_i$, and the second half to $T'_i$. Denote by $T' := \dot\cup_{i \in \{1,\ldots,l\}} T'_i$; then $R' : \Sigma \rightarrow T \dot\cup T'$ further divides the agents in each group $i$ into two subgroups corresponding to $T_i$ and $T'_i$, respectively. Extension that divides a group into three or more subgroups follows similarly.  Under the new relabeling map $R'$, there are two distinct template generators for each group $i$:
\begin{align*}
& R'({\bf G}_{ij}) = {\bf H}_i,\ j \in [1, \lfloor \frac{n_i}{2} \rfloor]\\
& R'({\bf G}_{ij}) = {\bf H}'_i,\ j \in [\lfloor \frac{n_i}{2} \rfloor+1, n_i].
\end{align*}
Consider the following extension of (P1):

(P1$'$):  Compute the relabeled plant ${\bf H}'$ as the synchronous product of the generators ${\bf H}_i || {\bf H}'_i$, i.e.
\begin{flushright}
${\bf H}' := ||_{i \in \{1,...,l\}} ({\bf H}_i || {\bf H}'_i).$ \ \ \ \ \ \ \ \ \ \ \ \ \ \ \ \ \ (5)
\end{flushright}
As so constructed, ${\bf H}'$ allows at most two agents in the same group to work in parallel.
Proceed with the same (P2)-(P4) as in Section III, and denote the resulting supervisor by ${\bf SSUP}'$. The state size of ${\bf SSUP}'$ and its computation do not depend on the number of component agents, but depend on (5) and the number of groups. We have the following result.

\begin{Proposition} \label{prop:rest2}
Consider the plant {\bf G} as described in Section~II.B and suppose that Assumptions (A1), (A2), (A3), and (A4) hold.
If $L_m(\textbf{H}')$ is controllable with respect to $R'(L(\textbf{G}))$, then ${\bf SSUP}'$ is a scalable supervisor that solves SSCSP.
\end{Proposition}

The proof of Proposition~\ref{prop:rest2} is similar to that of Theorem 1, by replacing $R$, {\bf H} and {\bf SSUP} by $R'$,
${\bf H}'$ and ${\bf SSUP'}$ throughout. The condition of Proposition~\ref{prop:rest2} is efficiently checkable, due analogously to Propositions~1 and 2.

Note that the improved permissiveness comes at a cost of increased computation effort.
The more subgroups are divided by `refining' the relabeling map,
the more agents in the same group are allowed to work in parallel in the relabeled plant ${\bf H}'$, and the more computational cost for deriving ${\bf SSUP'}$. In Section VI. A below, we shall illustrate the  method presented in this section by an example.

\section{Scalable Distributed Control} \label{sec4_sloc}

So far we have synthesized a scalable supervisor {\bf SSUP} that effectively controls the entire multi-agent system, i.e. {\bf SSUP} is a {\it centralized} controller. For the type of system considered in this paper which consists of many independent agents, however, it is also natural to design a {\it distributed} control architecture where each individual agent acquires its own local controller (thereby becoming autonomous)\footnote{In the centralized architecture, the communication from {\bf SSUP} to the agents is typically done via event broadcasting. On the other hand, in a distributed architecture, the communication between local controllers of the agents is naturally pairwise.}.

Generally speaking, a distributed control architecture is advantageous in reducing (global) communication load, since local controllers typically need to interact only with their (nearest) neighbors. A distributed architecture might also be more fault-tolerant, as partial failure of local controllers or the corresponding agents would unlikely to overhaul the whole system.

For these potential benefits, we aim in this section to design for the multi-agent system a distributed control architecture.
In particular, we aim to design local controllers that have the same {\it scalability} as the centralized {\bf SSUP}; namely their state sizes and computation are independent of the number of agents in the system. Thus when some agents break down and/or new agents are added in, there is no need of recomputing or reconfiguring these local controllers.

Let us now formulate the following Scalable Distributed Control Synthesis Problem (SDCSP):

\smallskip
{\it
Design a set of scalable local controllers ${\bf SLOC}_{ij}$ (a nonblocking generator), one for each agent ${\bf G}_{ij}$ ($i\in \{1,...,l\}$, $j \in \{1,...,n_i\}$) such that

\noindent (i) The number of states and computation of ${\bf SLOC}_{ij}$ are independent of the number $n_i$ of agents for all $i \in \{1,\ldots,l\}$;

\noindent (ii) the set of ${\bf SLOC}_{ij}$ is (collectively) {\it control equivalent} to the scalable supervisor ${\bf SSUP}$ with respect to plant ${\bf G}$, i.e.
{\small \begin{align} \label{eq:sloc_problem}
\left( \bigcap_{\substack{i\in \{1,...,l\} \\ j \in \{1,...,n_i\}}} L_m({\bf SLOC}_{ij}) \right) \cap L_m({\bf G})= L_m({\bf SSUP}) \cap L_m({\bf G}).
\end{align}
}}
\smallskip

To solve SDCSP, we employ a known technique called {\it supervisor localization} \cite{Cai & Wonham (2010),Cai & Wonham (2015),Cai & Wonham (2016)}, which works to decompose an arbitrary supervisor into a set of local controllers whose collective behavior is equivalent to that supervisor. Since we have synthesized {\bf SSUP}, the scalable supervisor, a straightforward approach would be to apply supervisor localization to decompose the associated controlled behavior $L_m(\textbf{SSUP})\cap L_m(\textbf{G})$.\protect\footnote[7]{Note that it is incorrect to localize $L_m(\textbf{SSUP})$, because $L_m(\textbf{SSUP})$ is in general not controllable with respect to $L({\bf G})$.} This approach would require, however, the computation of {\bf G} which is infeasible for large systems and cause the resulting local controllers non-scalable.

Instead we propose the following procedure for designing scalable local controllers $\textbf{SLOC}_{ij}$, for $i\in \{1,...,l\}$ and $j \in \{1,...,n_i\}$.

\noindent (Q1) Apply supervisor localization to decompose the relabeled supervisor {\bf RSUP} into {\it relabeled local controllers} ${\bf RLOC}_{i}$, $i\in \{1,...,l\}$, such that \cite{Cai & Wonham (2016)}
\begin{align*}
\left( \bigcap_{\substack{i\in \{1,...,l\}}} L_m({\bf RLOC}_{i}) \right) \cap L_m({\bf H})= L_m({\bf RSUP}).
\end{align*}

\noindent (Q2) Compute $\mbox{trim}( \textbf{RLOC}_i \| \textbf{H}_i )$, where trim($\cdot$) operation removes blocking states (if any) of the argument generator.


\noindent (Q3) Inverse-relabel $\mbox{trim}( \textbf{RLOC}_i \| \textbf{H}_i )$ to obtain $\textbf{SLOC}_{ij}$ ($j \in \{1,...,n_i\}$), i.e.
\begin{align} \label{eq:sloc}
\textbf{SLOC}_{ij} := R^{-1}( \mbox{trim}( (\textbf{RLOC}_i \| \textbf{H}_i) ) ).
\end{align}

\medskip

Notice that the computations involved in the above procedure are independent of the number $n_i$ ($i\in \{1,...,l\}$) of agents.
In (Q1), computing ${\bf RLOC}_{i}$ by localization requires computing {\bf RSUP} and {\bf H} (in (P1) and (P3) respectively), both of which are independent of $n_i$.
In (Q2), for the synchronous product both ${\bf RLOC}_{i}$ and ${\bf H}_i$ are independent of $n_i$, while trim may only reduce some states.
Finally in (Q3), inverse-relabeling does not change the number of states. Therefore the state number of the resulting scalable local controller $\textbf{SLOC}_{ij}$ and its computation are independent of the number $n_i$ ($i\in \{1,...,l\}$) of agents.

The synchronous product in (Q2) is indeed crucial to ensure the correctness of the resulting local controllers.
If we did not compute this synchronous product and set the local controllers to be $R^{-1}( \mbox{trim}( \textbf{RLOC}_i ) )$, then
such local controllers cannot even guarantee that the controlled behavior satisfies the imposed specification, as will be demonstrated in Section~V.A below.

On the other hand, the synchronous product in (Q2) may produce blocking states; such an example is provided in Section~V.B.
Thus the trim operation is needed to ensure that the resulting ${\bf SLOC}_{ij}$ is a nonblocking generator.

In addition, note that $\textbf{SLOC}_{ij}$ are the same for all $j \in \{1,...,n_i\}$. This means that every agent ${\bf G}_{ij}$ in the same group $\mathcal{G}_i$ obtains the same local controller, although each local controller will be dedicated to enabling/disabling only the controllable events originated from its associated agent.

The main result of this section is the following.


\begin{Theorem} \label{thm:local}
The set of ${\bf SLOC}_{ij}$ ($i\in \{1,...,l\}$, $j \in \{1,...,n_i\}$) as in (\ref{eq:sloc}) is a set of scalable local controllers that solves SDCSP.
\end{Theorem}

{\it Proof:} That the number of states of ${\bf SLOC}_{ij}$ and its computation are independent of the number $n_i$ of agents for all $i\in \{1,\ldots,l\}$, $j \in \{1,...,n_i\}$ has been asserted following (Q3) of designing ${\bf SLOC}_{ij}$. Hence to prove that the set of ${\bf SLOC}_{ij}$ is a set of scalable local controllers that solves SDCSP, we will show (\ref{eq:sloc_problem}).

From (Q1) we have
{\small \begin{align*}
&\left( \bigcap_{\substack{i\in \{1,...,l\}}} L_m({\bf RLOC}_{i}) \right) \cap L_m({\bf H})= L_m({\bf RSUP}) \\
\Rightarrow &\left( \bigcap_{\substack{i\in \{1,...,l\}}} L_m({\bf RLOC}_{i}) \right) \cap \left( \|_{\substack{i\in \{1,...,l\}}} L_m({\bf H}_{i}) \right)= L_m({\bf RSUP}) \\
\Rightarrow &\bigcap_{\substack{i\in \{1,...,l\}}} \left( L_m({\bf RLOC}_{i}) \| L_m({\bf H}_{i}) \right)= L_m({\bf RSUP}) \\
\Rightarrow &\bigcap_{\substack{i\in \{1,...,l\}}} L_m( \mbox{trim}( {\bf RLOC}_{i} \| {\bf H}_{i}) ) = L_m({\bf RSUP})
\end{align*} }
Inverse-relabeling both sides and applying Lemma~\ref{lem:cr}(iv), we derive
{\small \begin{align*}
&R^{-1}\left( \bigcap_{\substack{i\in \{1,...,l\}}} L_m( \mbox{trim}( {\bf RLOC}_{i} \| {\bf H}_{i}) ) \right)= R^{-1}(L_m({\bf RSUP})) \\
\Rightarrow &\bigcap_{\substack{i\in \{1,...,l\}}} R^{-1}( L_m( \mbox{trim}( {\bf RLOC}_{i} \| {\bf H}_{i}) ) )= R^{-1}(L_m({\bf RSUP})) \\
\Rightarrow &\bigcap_{\substack{i\in \{1,...,l\}}} L_m( R^{-1}(\mbox{trim}( {\bf RLOC}_{i} \| {\bf H}_{i}) ) )= R^{-1}(L_m({\bf RSUP}))
\end{align*} }
Finally it follows from (\ref{eq:sloc}) and (P4) that
{\small \begin{align*}
&\left( \bigcap_{\substack{i\in \{1,...,l\} \\ j \in \{1,...,n_i\}}} L_m({\bf SLOC}_{ij}) \right)= L_m({\bf SSUP})\\
\Rightarrow &\left( \bigcap_{\substack{i\in \{1,...,l\} \\ j \in \{1,...,n_i\}}} L_m({\bf SLOC}_{ij}) \right) \cap L_m({\bf G})= L_m({\bf SSUP}) \cap L_m({\bf G}).
\end{align*} }
That is, (\ref{eq:sloc_problem}) is established. \qed

\section{Illustrating Examples}

In this section, we provide three examples to illustrate our proposed scalable supervisory synthesis as well as distributed control.
The first example is the extension of the small factory example (studied in Figs.~\ref{fig:SmallFact22}, \ref{fig:SmallFact22_ScaSup}) to arbitrary numbers of input and output machines.  The second example is a transfer line system, where we illustrate how to deal with more than one specification. The last example is called mutual exclusion, where the plant naturally contains only one group of agents; we demonstrate how to fit this type of multi-agent systems into our setting and apply our method to derive scalable supervisors and local controllers.

\subsection{Small Factory}

This example has already been presented in Figs.~\ref{fig:SmallFact22} and \ref{fig:SmallFact22_ScaSup}, with 3 input machines and 2 output machines. Here we consider the general case where there are $n$ input machines and $m$ output machines, for arbitrary $n,m \geq 1$. 
\begin{figure}[t!]
  \centering
  \includegraphics[width=0.28\textwidth]{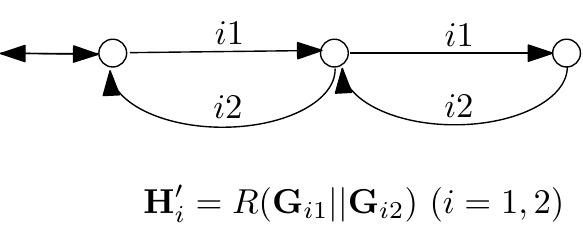}
  \caption{${\bf H}'_i$ using method in Subsection IV.A.}
  \label{fig:HN}
\end{figure}

\begin{figure}[t!]
  \centering
  \includegraphics[width=0.5\textwidth]{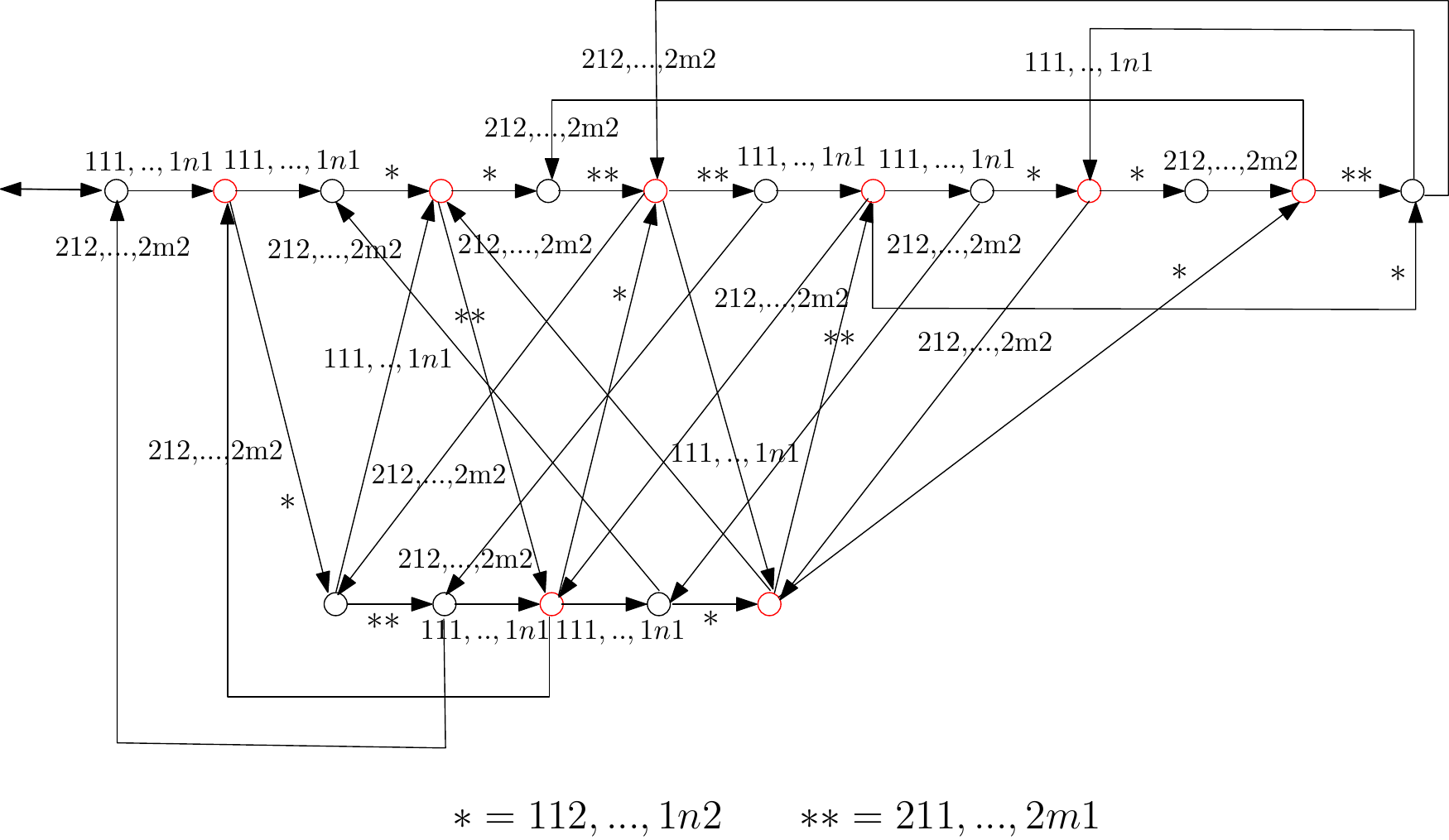}
  \caption{Small Factory: maximally permissive and scalable supervisor ${\bf SSUP'}$ derived using methods in Section IV. }
  \label{fig:SUPHN}
\end{figure}

To improve permissiveness of {\bf SSUP}, we employ the method presented in Section IV. 
Define a new relabeling map as follows:
\begin{align*}
& R'(1j1) = 11, R'(1j2) = 12,\ \mbox{for}\ j \in [1, \lfloor \frac{n}{2} \rfloor]\\
& R'(1j1) = 11', R'(1j2) = 12',\ \mbox{for}\ j \in [\lfloor \frac{n}{2} \rfloor +1, n]\\
& R'(2j1) = 21, R'(2j2) = 22,\ \mbox{for}\ j \in [1, \lfloor \frac{m}{2} \rfloor]\\
& R'(2j1) = 21', R'(2j2) = 22',\ \mbox{for}\ j \in [\lfloor \frac{m}{2} \rfloor +1, m].\\
\end{align*}
Accordingly there are two distinct template generators for each of the two groups:
\begin{align*}
& R'({\bf G}_{1j}) = {\bf H}_1,\ j \in [1, \lfloor \frac{n}{2} \rfloor]\\
& R'({\bf G}_{1j}) = {\bf H}'_1,\ j \in [\lfloor \frac{n}{2} \rfloor+1, n]\\
& R'({\bf G}_{2j}) = {\bf H}_2,\ j \in [1, \lfloor \frac{m}{2} \rfloor]\\
& R'({\bf G}_{2j}) = {\bf H}'_2,\ j \in [\lfloor \frac{m}{2} \rfloor+1, m].\\
\end{align*}
Compute the relabeled plant ${\bf H}' :=  ({\bf H}_1 || {\bf H}'_1) || ({\bf H}_2 || {\bf H}'_2)$. Proceeding with the same (P2)-(P4) as in Section III, we derive again the ${\bf SSUP}'$ in Fig.~\ref{fig:SUPHN}. Therefore, the method in Section IV lead to a more permissive scalable supervisor (at the cost of increased computational effort), in this particular example maximally permissive.

\subsection{Transfer Line}
\begin{figure}[t!]
\centering
  \includegraphics[width=0.5\textwidth]{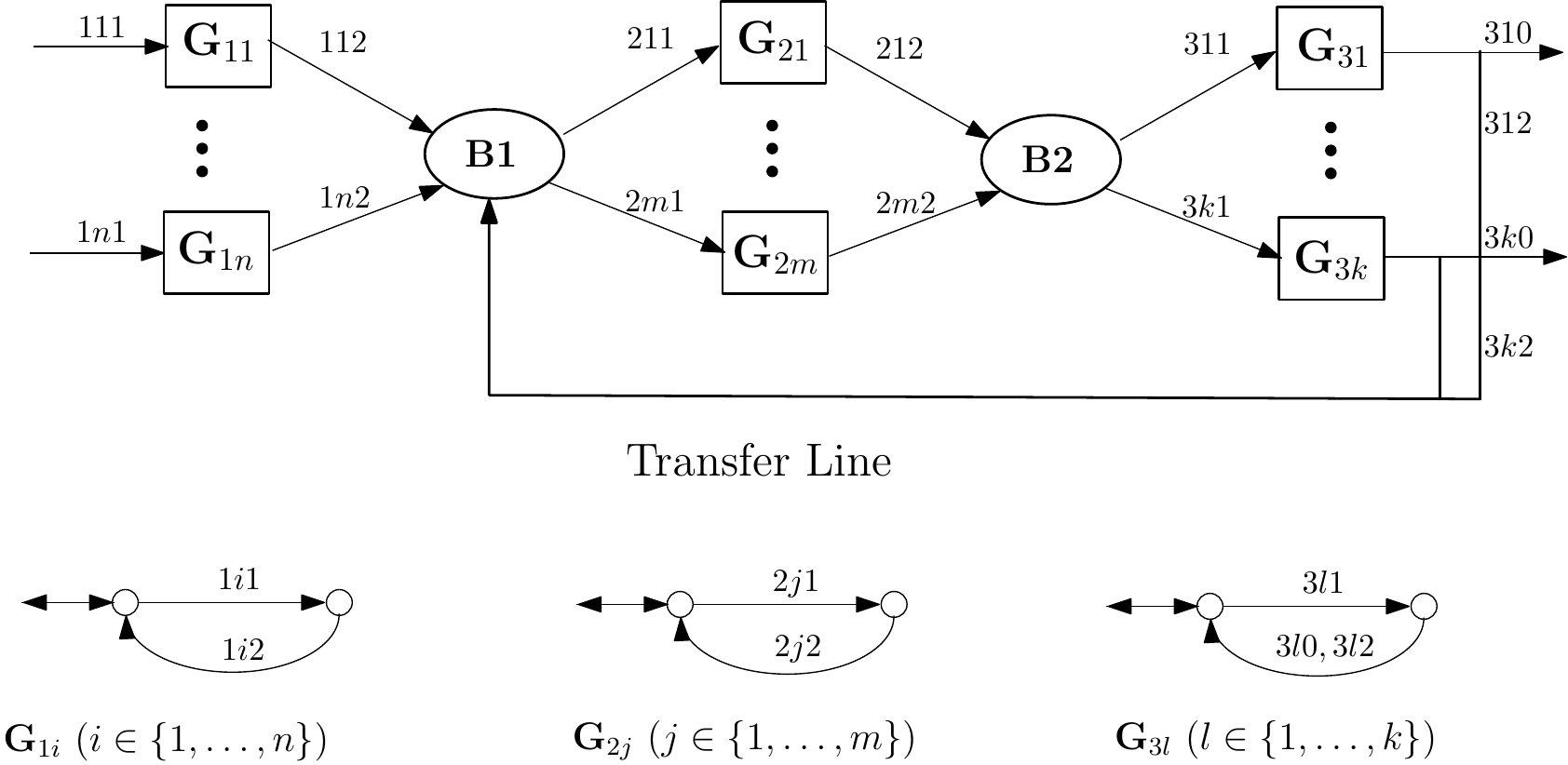}\\
  \caption{Transfer line: system configuration and component agents. Event $1i1$ ($i \in \{1,...,n\}$) means that ${\bf G}_{1i}$ starts to work by taking in a workpiece, and $1i2$ means that ${\bf G}_{1i}$ finishes work and deposits a workpiece to buffer {\bf B1}; event $2j1$ ($j \in \{1,...,m\}$) means that ${\bf G}_{2j}$ starts to work by taking in a workpiece, and $2j2$ means that ${\bf G}_{2j}$ finishes work and deposits a workpiece to buffer {\bf B2}; event $3l1$ ($l \in \{1,...,k\}$) means that ${\bf G}_{3l}$ starts to work by testing a workpiece, $3l0$ means that ${\bf G}_{3l}$ detects a fault and sends the faulty workpiece back to buffer {\bf B1}, and $3l2$ means that ${\bf G}_{3l}$ detects no fault and output the successfully processed workpiece.}
  \label{fig:TL}
\end{figure}

The second example we present is a transfer line system, adapted from \cite{Wonham (2016)}. In this example, we demonstrate how to deal with the case where the overall specification is composed from two independent ones. As displayed in Fig.~\ref{fig:TL}, transfer line consists of machines (${\bf G}_{11},\ldots,{\bf G}_{1n}$; ${\bf G}_{21},\ldots,{\bf G}_{2m}$) and test units (${\bf G}_{31},\ldots,{\bf G}_{3k}$), linked by two buffers {\bf B1} and {\bf B2} both with capacities 1.
The generators of the agents are shown in Fig.~\ref{fig:TL}. Based on their different roles, the machines are divided into 3 groups:
\begin{center}
$\mathcal{G}_1=\{\textbf{G}_{11},\ldots,\textbf{G}_{1n}\}$

$\mathcal{G}_2=\{\textbf{G}_{21},\ldots,\textbf{G}_{2m}\}$

$\mathcal{G}_3=\{\textbf{G}_{31},\ldots,\textbf{G}_{3k}\}$.
\end{center}
Let the relabeling map $R$ be given by
\begin{align*}
& R(1i1)=11,\ R(1i2)=12,\ i \in \{1,\ldots,n\} \\
& R(2j1)=21,\ R(2j2)=22,\ j \in \{1,\ldots,m\} \\
& R(3l0)=30,\ R(3l1)=31,\ R(3l2)=32,\ l \in \{1,\ldots,k\}
\end{align*}
where odd-number events are controllable and even-number events are uncontrollable.
It is easily observed that Assumptions (A1), (A2) hold.

\begin{figure}[t!]
  \centering
  \includegraphics[width=0.35\textwidth]{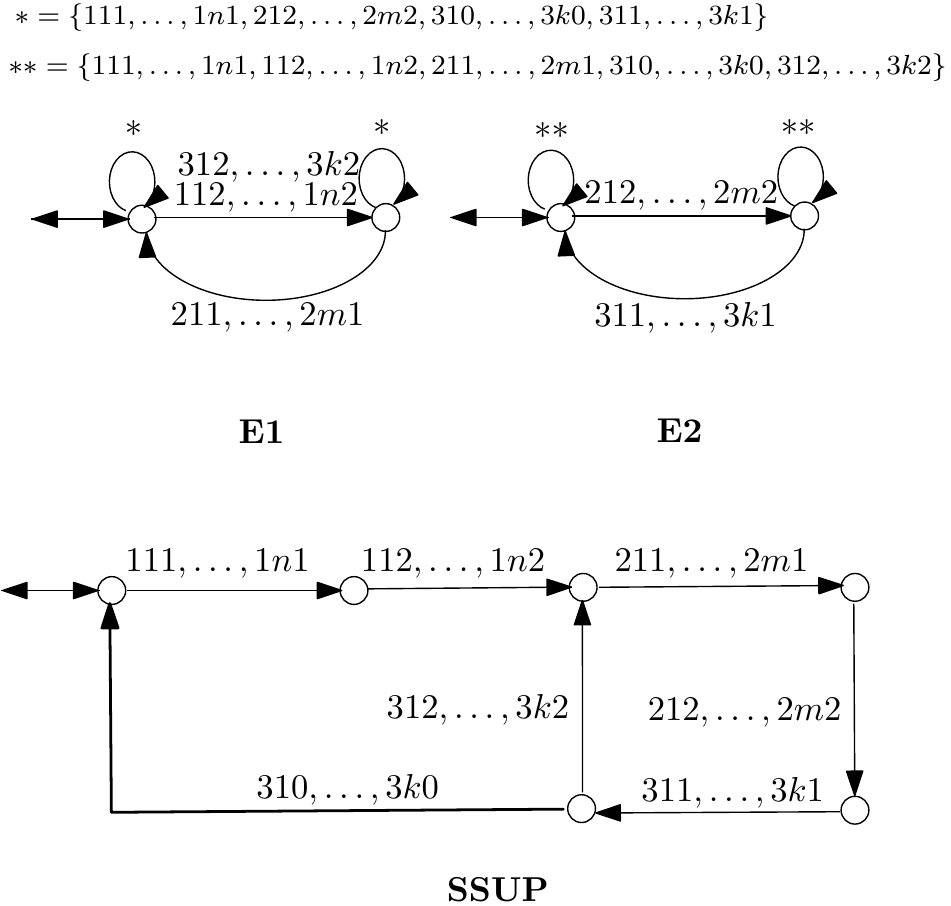}\\
  \caption{Transfer line: specification generators {\bf E1}, {\bf E2}, and scalable supervisor {\bf SSUP}}
  \label{fig:TL_SSUP}
\end{figure}

The specification is to avoid underflow and overflow of buffers {\bf B1} and {\bf B2}, which is enforced by the two generators {\bf E1} and {\bf E2} in Fig.~\ref{fig:TL_SSUP}. Thus the overall specification $E$ is $E = L_m({\bf E1}) \cap L_m({\bf E2})$, which can be verified to satisfy Assumption (A3). It is also verified that  Assumption (A4) holds.
In addition, it is checked that $L_m({\bf H}_i) := L_m(R(\textbf{G}_{i1}))$ ($i=1,2,3$) is controllable with respect to $R(L(\textbf{G}_{i1}||\textbf{G}_{i2}))$. By Proposition~\ref{prop:check}, we have that $L_m({\bf M})$ is controllable with respect to $R(L({\bf G}))$. Therefore the sufficient condition of Theorem~1 is satisfied.

By the procedure (P1)-(P4)  with $k_1=2,\ k_2=3,\ k_3=1$, we design a scalable supervisor {\bf SSUP}, displayed in Fig.~\ref{fig:TL_SSUP}. The state size of {\bf SSUP} and its computation are independent of the agent numbers $n,m,k$. Moreover, the controlled behavior of {\bf SSUP} is in fact equivalent to that of the monolithic supervisor {\bf SUP}, i.e. $L_m({\bf SSUP}) \cap L_m(\textbf{G}) = L_m({\bf SUP})$, for arbitrary fixed values of $n,m,k$. This is owing to that both buffers have only one slot, and thus the restriction due to relabeling is already enforced by the monolithic supervisor in order to satisfy the specification.
\begin{figure}[t!]
  \centering
  \includegraphics[width=0.48\textwidth]{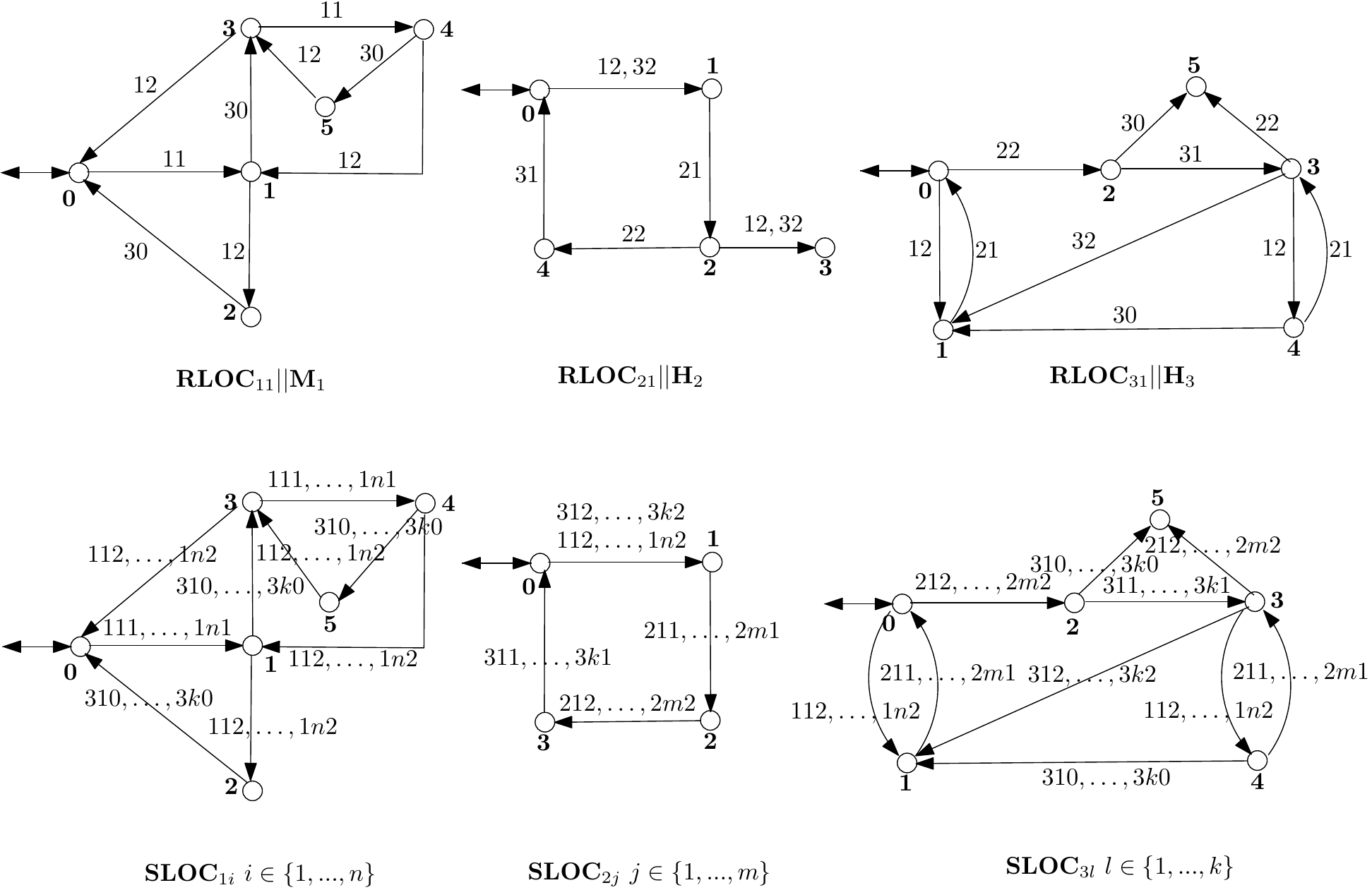}\\
  \caption{Transfer line: scalable local controllers ($\textbf{SLOC}_{1i}$ for machine ${\bf G}_{1i}$, $i \in \{1,...,n\}$; $\textbf{SLOC}_{2i}$ for machine ${\bf G}_{2j}$, $j \in \{1,...,m\}$; $\textbf{SLOC}_{3i}$ for test unit ${\bf G}_{3l}$, $l \in \{1,...,k\}$)}
  \label{fig:TLSLOC}
\end{figure}

{\bf Scalable distributed control.} Following the procedure (Q1)-(Q3) in Section~4, we compute the scalable local controllers for the individual agents.
In (Q2), certain synchronous products turn out to be blocking, as displayed in Fig.~\ref{fig:TLSLOC} (upper part). Hence the trim operation in (Q2) is important to ensure that the resulting local controllers are nonblocking.
In Fig.~\ref{fig:TLSLOC} (lower part), $\textbf{SLOC}_{1i}$ (6 states) is for the machine ${\bf G}_{1i}$, $i \in \{1,...,n\}$; $\textbf{SLOC}_{2j}$ (4 states) for the machine ${\bf G}_{2j}$, $j \in \{1,...,m\}$; and $\textbf{SLOC}_{3i}$ (6 states) for the test unit ${\bf G}_{3l}$, $l \in \{1,...,k\}$. It is verified that the desired control equivalence between the set of local controllers and the supervisor {\bf SSUP} in Fig.~\ref{fig:TL_SSUP} is satisfied, i.e. the condition (ii) of SDCSP holds.

The control logic of the scalable local controllers is as follows. First for $\textbf{SLOC}_{1i}$ ($i \in \{1,...,n\}$), which controls only the event $1i1$ of machine ${\bf G}_{1i}$, observe that event $1i1$ is disabled at states 1, 2, and 4 to protect buffer {\bf B1} against overflow, while it is disabled at states 5 due to the restriction of relabeling. As mentioned above, relabeling allows parallel operations of two machines in group one.

Next for $\textbf{SLOC}_{2j}$ ($j \in \{1,...,m\}$), which is responsible only for event $2j1$ of machine ${\bf G}_{2j}$, observe that event $2j1$ is disabled at states 0, 2 and 3. This is to protect buffer {\bf B1} against underflow and buffer {\bf B2} against overflow.

Finally for $\textbf{SLOC}_{3l}$ ($l \in \{1,...,k\}$), which is responsible only for event $3l1$ of test unit ${\bf G}_{3l}$, observe that event $3l1$ is disabled at states 0, 1, 3, 4 and 5. This is to protect buffer {\bf B2} against underflow and buffer {\bf B1} against overflow.

\subsection{Mutual Exclusion}

In this last example, mutual exclusion, we demonstrate how to transform the problem into our setup and apply our scalable supervisory synthesis.  There are $n (>1)$ agents that compete to use a single resource; the specification is to prevent the resource being simultaneously used by more than one agent.
\begin{figure}[t!]
  \centering
   \includegraphics[width=0.45\textwidth]{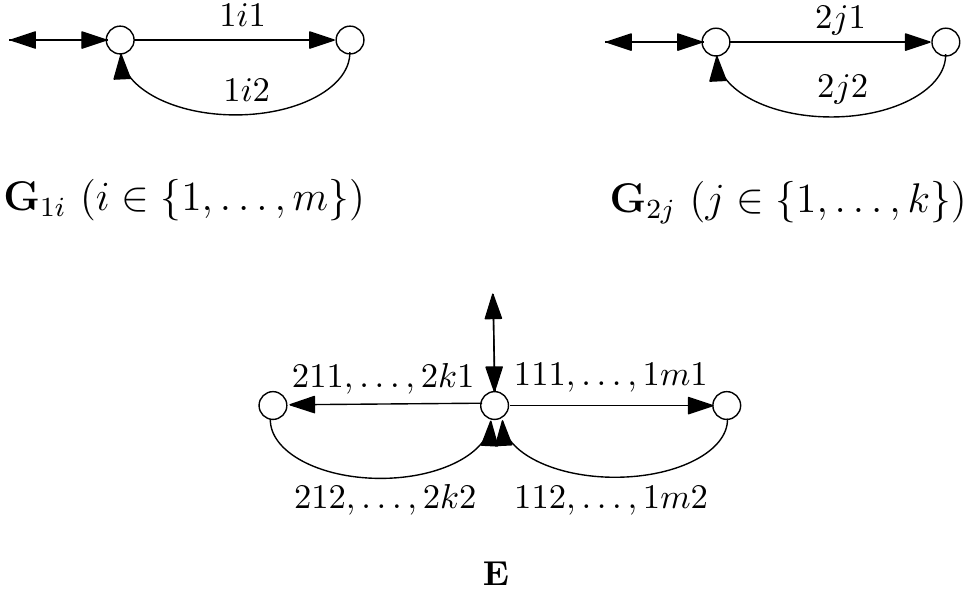}\\
  \caption{Mutual exclusion: component agents and specification. Event $1i1$ ($i \in \{1,...,m\}$) means that ${\bf G}_{1i}$ starts using the resource,  $1i2$ means that ${\bf G}_{1i}$ finishes using the resource; event $2j1$ ($j \in \{1,...,k\}$) means that ${\bf G}_{2j}$ starts using the resource,  $2j2$ means that ${\bf G}_{2j}$ finishes using the resource.}
  \label{fig:MX}
\end{figure}

For this problem, it is natural to treat all agents as just one group. However, our approach would then relabel every agent to a single template model, to which the mutual exclusion specification could not be imposed (mutual exclusion specifies requirement {\it between} different agents).
Thus in order to apply our synthesis method, we (artificially) separate the agents into two groups, with $m$ and $k$ agents respectively, such that $n=m+k$. Namely
\begin{center}
$\mathcal{G}_1=\{\textbf{G}_{11},\ldots,\textbf{G}_{1m}\}$

$\mathcal{G}_2=\{\textbf{G}_{21},\ldots,\textbf{G}_{2k}\}$.
\end{center}
The generators of the agents separated into two groups and the specification are displayed in Fig.~\ref{fig:MX}.

Let the relabeling map $R$ be given by
\begin{align*}
& R(1i1)=11,\ R(1i2)=12,\ i \in \{1,\ldots,m\} \\
& R(2j1)=21,\ R(2j2)=22,\ j \in \{1,\ldots,k\}
\end{align*}
where odd-number events are controllable and even-number events are uncontrollable.
It is readily checked that Assumptions (A1), (A2) hold. Moreover, it is verified that ${\bf H}_i := L_m(R(\textbf{G}_{i1}))$ ($i=1,2$) is controllable with respect to $R(L(\textbf{G}_{i1}||\textbf{G}_{i2}))$, and $R^{-1}R({\bf E})={\bf E}$; hence the sufficient condition of Theorem~1 is satisfied.

\begin{figure}[t!]
  \centering
  \includegraphics[width=0.25\textwidth]{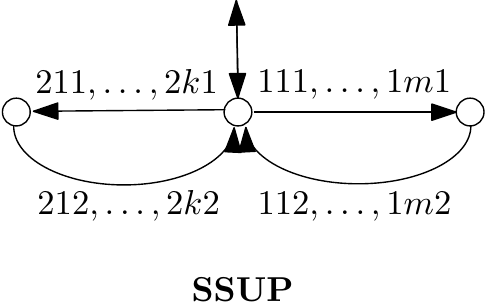}
  \caption{Mutual exclusion: scalable supervisor {\bf SSUP}}
  \label{fig:MX_SSUP}
\end{figure}

By the procedure (P1)-(P4) with $k_1=1,\ k_2=1$, we design a scalable supervisor {\bf SSUP}, displayed in Fig.~\ref{fig:MX_SSUP}. Note that {\bf SSUP} is identical to the specification {\bf E}, and the state size of {\bf SSUP} and its computation are independent of the agent numbers $m,k$ (hence the total number $n$). Moreover, the controlled behavior of {\bf SSUP} is equivalent to that of the monolithic supervisor {\bf SUP}, i.e. $L_m({\bf SSUP}) \cap L_m(\textbf{G}) = L_m({\bf SUP})$, for any fixed value of $n$. This is because there is only a single resource, and no matter how many agents are in the system, the resource can be used by only one agent at any given time. Thus the restriction due to relabeling has already been imposed by the mutual exclusion specification and enforced by the monolithic supervisor {\bf SUP}.

\begin{figure}[t!]
  \centering
  \includegraphics[width=0.35\textwidth]{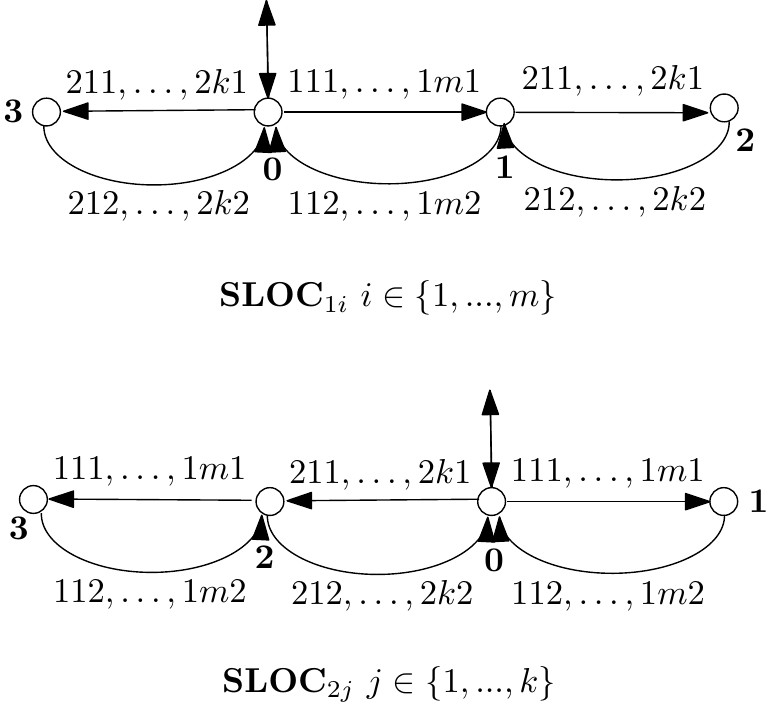}\\
  \caption{Scalable local controllers for mutual exclusion:  $\textbf{SLOC}_{1i1}$ and $\textbf{SLOC}_{2j1}$ are the scalable local controllers of event $1i1$ and $2j1$ respectively; $R^{-1}(\textbf{RLOC}_{11})$ and $R^{-1}(\textbf{RLOC}_{21})$ are obtained by inverse relabeling the local controllers of event 11 and 21; $\textbf{LOC}_{1i1}$ and $\textbf{LOC}_{2j1}$ are calculated by scalable supervisor ${\bf SSUP}$ and plant ${\bf G}$.}
  \label{fig:MESLOC}
\end{figure}

{\bf Scalable distributed control.} Following the procedure (Q1)-(Q3) in Section~IV, we compute the scalable local controllers for the individual agents. Specifically, as displayed in Fig.~\ref{fig:MESLOC}, $\textbf{SLOC}_{1i}$ (4 states) is for the first-group agent ${\bf G}_{1i}$, $i \in \{1,...,m\}$; while $\textbf{SLOC}_{2j}$ (4 states) is for the second-group agent ${\bf G}_{2j}$, $j \in \{1,...,k\}$. It is verified that the desired control equivalence between the set of local controllers and the supervisor {\bf SSUP} in Fig.~\ref{fig:MX_SSUP} is satisfied, i.e. (\ref{eq:sloc_problem}) holds.

The control logic of the scalable local controllers is as follows. First for $\textbf{SLOC}_{1i}$ ($i \in \{1,...,m\}$), which controls only the event $1i1$ of the first-group agent ${\bf G}_{1i}$, observe that event $1i1$ is disabled at states 1, 2, and 3. At all these states, the resource is being used by some agent; hence by mutual exclusion event $1i1$ must be disabled.

It is worth noting that if the sequence $1i1.2j1$ ($j \in \{1,...,k\}$) occurred, which is allowed by $\textbf{SLOC}_{1i}$, the mutual exclusion specification would be violated. Indeed $2j1$ must be disabled after the occurrence of $1i1$. However, since the local controller $\textbf{SLOC}_{1i}$ is responsible only for event $1i1$, the correct disablement of $2j1$ ($j \in \{1,...,k\}$) is left for another dedicated local controller $\textbf{SLOC}_{2j}$. As we can see in $\textbf{SLOC}_{2j}$, event $2j1$ is disabled at states 1, 2, and 3. In particular, at state 1 (i.e. after $1i1$ occurs) event $2j1$ is correctly disabled to guarantee mutual exclusion (as expected). Therefore, while each local controller enables/disables only its locally-owned events, together they achieve correct global controlled behavior.

\section{CONCLUSIONS}

We have studied multi-agent discrete-event systems that can be divided into several groups of independent and similar agents. We have employed a relabeling map to generate template structures, based on which scalable supervisors are designed whose state sizes and computational process are independent of the number of agents. We have presented a sufficient condition for the validity of the designed scalable supervisors, and shown that this condition may be verified with low computational effort. Moreover, based on the scalable supervisor we have designed scalable local controllers, one for each component agent.
Three examples have been provided to illustrate our proposed synthesis methods.

In future research, we aim to find conditions under which scalable supervisors may be designed to achieve controlled behavior identical to the monolithic supervisor.  We also aim to search for new designs of scalable supervisors when the sufficient condition of Theorem~1 fails to hold.
Additionally we are interested in investigating, in the context of scalable supervisory control, the issue of partial observation.




\section{Appendix}

{\it Proof of Lemma~\ref{lem:cr}:}

(i): ($\subseteq$) Let $t \in R(\overline{L})$. There exists $s\in \overline{L}$ such that $t=R(s)$. Let $w\in \Sigma^{*}$ be such that $sw \in L$. Hence $R(sw)=R(s)R(w) \in R(L)$. Therefore $R(s) \in \overline{R(L)}$, i.e. $t \in \overline{R(L)}$.

($\supseteq$) Let $t \in \overline{R(L)}$. There exists $u \in T^{*}$ such that $tu \in R(L)$. Thus there are strings $s,w \in \Sigma^{*}$ such that $sw \in L$, $R(s)=t$ and $R(v)=w$. Therefore $s\in \overline{L}$, so $R(s)\in R(\overline{L})$, i.e. $t \in R(\overline{L})$.

\medskip

(ii): Let $t \in R(L_1\cap L_2)$. There exists $s\in L_1\cap L_2$ such that $R(s)=t$. Thus $s\in L_1$ and $s\in L_2$. It follows that $R(s) \in R(L_1)$ and $R(s) \in R(L_2)$, i.e. $t = R(s) \in R(L_1)\cap R(L_2)$.

\medskip

(iii): ($\subseteq$) Let $s \in R^{-1}(\overline{H})$. Then $R(s)\in \overline{H}$, and thus there exists $R(t)\in T^{*}$ such that $R(s)R(t)\in H$, i.e. $R(st)\in H$. It follows that $st \in R^{-1}(H)$, therefore $s \in \overline{R^{-1}(H)}$.

($\supseteq$) Let $s \in \overline{R^{-1}(H)}$. Then there exists $t\in \Sigma^{*}$ such that $st \in R^{-1}(H)$; so $R(st)\in H$, i.e. $R(s)R(t)\in H$. Thus $R(s)\in \overline{H}$, and therefore $s \in R^{-1}(\overline{H})$.

\medskip

(iv): ($\subseteq$) Let $s \in R^{-1}(H_1 \cap H_2)$. Then $R(s)\in H_1\cap H_2$, i.e. $R(s)\in H_1$ and $R(s)\in H_2$. Hence $s \in R^{-1}(H_1)$ and $s \in R^{-1}(H_2)$, i.e. $s \in R^{-1}(H_1) \cap R^{-1}(H_2)$.

($\supseteq$) Let $s \in R^{-1}(H_1)\cap R^{-1}(H_2)$, i.e. $s \in R^{-1}(H_1)$ and $s \in R^{-1}(H_2)$.
Thus $R(s)\in H_1$ and $R(s)\in H_2$, i.e. $R(s)\in H_1\cap H_2$. Therefore $s \in R^{-1}(H_1\cap H_2)$. \qed

{\it Proof of Lemma~\ref{lem:sub}:} First, we show that {\bf H} is nonblocking. By Assumption~(A1) each plant component ${\bf G}_{ij}$ is nonblocking. Thus by (\ref{eq:similarset}) and Lemma~\ref{lem:nonb}, each ${\bf H}_i$ is also nonblocking. Therefore by Assumption~(A2) that ${\bf H}_i$ do not share events, we derive that {\bf H} computed as the synchronous product of ${\bf H}_i$ is nonblocking.

Next, we prove $L_m({\bf H}) \subseteq R(L_m({\bf G}))$. From (P1) we have
\begin{align*}
L_m(\textbf{H})&=||_{i \in \{1,...,l\}}L_m(\textbf{H}_i)\\
&=||_{i \in \{1,...,l\}} R(L_m(\textbf{G}_{i1})) \ \ \ \ \ \mbox{(by (\ref{eq:Hi}))} \\
&\subseteq ||_{i \in \{1,...,l\}} \left( R( L_m(\textbf{G}_{i1}) || \cdots || L_m(\textbf{G}_{i \, n_i})) \right)\\
&= R\left( ||_{i \in \{1,...,l\}}  (L_m(\textbf{G}_{i1}) || \cdots || L_m(\textbf{G}_{i \, n_i})) \right)\\
&\hspace{2.5cm} \mbox{(by Assumptions~(A1), (A2))}\\
&= R(L_m({\bf G})).
\end{align*}
\qed

\medskip

Finally we provide the proof of Lemma~\ref{lem:3.1}. This proof in fact has been given in the full version of \cite{Jiao & Gan (2015)}, which is currently under review and there is no online version we can refer to. For completeness (for the review of this paper), we reproduce the proof here.

\textit{Proof of Lemma~\ref{lem:3.1}:} $(\subseteq)$ This direction is always true.

$(\supseteq)$ Let $s\in \overline{L_m(\textbf{SSUP})}\cap \overline{L_m(\textbf{G})}$. Then
\begin{align*}
&s\in \overline{L_m(\textbf{SSUP})} \\
\Rightarrow &s\in \overline{R^{-1}(L_m(\textbf{RSUP}))} \ \ \ \mbox{(by (P4))}\\
\Rightarrow &s\in R^{-1}(\overline{L_m(\textbf{RSUP})}) \ \ \ \mbox{(by Lemma~\ref{lem:cr}(iii))}\\
\Rightarrow &R(s) \in \overline{L_m(\textbf{RSUP})}.
\end{align*}
Let $t:= R(s)$. Then there exists $w \in T^*$ such that $tw \in L_m({\bf RSUP})$.
By (P3) we have $L_m({\bf RSUP}) \subseteq L_m({\bf H})$, and by Lemma~\ref{lem:sub} $L_m({\bf H}) \subseteq R(L_m({\bf G}))$. Hence $tw \in R(L_m({\bf G}))$. This implies that there are $s'$ and $v'$ such that
\begin{align*}
R(s')=t \ \ \&\ \ R(v')=w \ \ \&\ \ s't' \in L_m({\bf G}).
\end{align*}
Since $R(s)=R(s')=t$ and by the symmetric structure of the plant under Assumptions (A1), (A2),
it can be shown that there exists $v$ such that $R(v) =R(v')=w$ and $sv \in L_m({\bf G})$.

On the other hand,
\begin{align*}
&tw \in L_m({\bf RSUP}) \\
\Rightarrow &R(s)R(v) \in L_m({\bf RSUP})\\
\Rightarrow &R(sv) \in L_m({\bf RSUP})\\
\Rightarrow &sv \in R^{-1}(L_m({\bf RSUP}))\\
\Rightarrow &sv \in L_m({\bf SSUP}).
\end{align*}
Hence
\begin{align*}
sv \in L_m({\bf SSUP}) \cap L_m({\bf G})
\end{align*}
by which we conclude that $s \in \overline{L_m({\bf SSUP}) \cap L_m({\bf G})}$. \qed

\end{document}